\documentclass[11pt, oneside]{article} 
\usepackage[utf8]{inputenc}
\usepackage[utf8]{inputenc}
\usepackage{geometry}                		
\usepackage{graphicx}				
\usepackage{amssymb}
\usepackage{amsmath}
\usepackage{subfig}
\usepackage{caption}
\usepackage{comment}
\usepackage{amsthm,verbatim,amsfonts,amscd,float,physics}
\usepackage{appendix}

\usepackage[utf8]{inputenc}
\usepackage{xcolor}
\numberwithin{equation}{section}
\usepackage{titling, lipsum}
\title{Quantum Computing of Schwarzschild-de Sitter Black Holes and Kantowski-Sachs Cosmology }
\author{Amy Joseph$^{(1)}$, Tristen White$^{(2)}$, Viti Chandra$^{(3)}$, Michael McGuigan$^{(4)}$\\
(1) Arizona State University (2) Florida State University (3) Half Hollow Hills High School \\
(4) email contact: michael.d.mcguigan@gmail.com
}
\date{}

\begin{document}
\begin{titlingpage}

\maketitle
\begin{abstract}
The quantum mechanics of Schwarzschild-de Sitter black holes is of great recent interest because of their peculiar thermodynamic properties  as well as their realization in modern dark energy cosmology which indicates the presence of a small positive cosmological constant.  We study Schwarzschild-de Sitter black holes and also the Kantowki-Sachs Cosmology using quantum computing. In these cases in addition to the Hamiltonian there is a Mass operator which plays an important role in describing the quantum states of the black hole and Kantowski-Sachs cosmology. We compute the spectrum of these operators using classical and quantum computing. For quantum computing we use the Variational Quantum Eigensolver which is hybrid classical-quantum algorithm that runs on near term quantum hardware. We perform our calculations using 4, 6 and 8 qubits in a harmonic oscillator basis, realizing the quantum operators of the Schwarzschild-de Sitter black hole and Kantowski-Sachs cosmology in terms of $16\times 16$, $64\times 64$ and $256 \times 256$ matrices respectively. For the 4 qubit case we find highly accurate results but for the other cases we find a more refined variational ansatz will be necessary to accurately represent the quantum states of a Schwarzschild-de Sitter black hole or Kantowki-Sachs cosmology accurately on a quantum computer.

\end{abstract}
\end{titlingpage}













\newpage 

\section{Introduction}

There is strong evidence that Universe is accelerating, has a positive cosmological constant and can be described by de Sitter space-time. Also there are direct observations of stellar and supermassive black holes which can be described by Schwarzschild or Kerr space-times. Thus one needs to consider the Schwarzschild-de Sitter space-time in order be consistent with these data. However conceptually this is a difficult case to consider quantum mechanically \cite{Shankaranarayanan:2003ya}
\cite{Susskind:2021omt}
\cite{Bousso:1997wi}
\cite{Bousso:2002fq}
\cite{Banks:2005bm}
\cite{Susskind:2021esx}
\cite{Giddings:2007nu}. For example the  AdS/CFT correspondence cannot be straightforwardly applied in this case. In addition models of quantum cosmology applied to, for example, the Kantowski-Sachs cosmology can suffer from difficulties with path integral quantization, such as the  Euclidean action not being bounded from below, which can cause  problems with  traditional Monte Carlo evaluation of path integrals. The Kantowski-Sachs path integral has similarities to the path integral for a black hole interior and thus is also relevant to Schwarzschild-de Sitter black holes \cite{Kantowski:1966te}
\cite{Fanaras:2021awm}
\cite{Halliwell:1990tu}
\cite{Maldacena:2019cbz}
\cite{Conradi:1994yy}
\cite{Pal:2015bma}
\cite{Pedram:2008wx}
\cite{Pedram:2010zzb}
\cite{Adamek:2010sg}
\cite{Kiefer:2008sw}
\cite{Wiltshire:1995vk}
\cite{Cavaglia:1996qp}
\cite{Kuchar:1994zk}
\cite{Kiefer:1998rr}. In this paper we study the application of quantum computing to the Schwarzschild-de Sitter black hole and the Kantowski-Sachs cosmology. Quantum computing has advantages for quantum simulation \cite{Joseph:2021naq}
\cite{Kocher:2018ilr}
\cite{Ganguly:2019kkm}
\cite{Liu:2020wtr}
\cite{Brown:2019hmk}
\cite{Nezami:2021yaq}
\cite{Bousso:2022ntt} and represents the Lorentzian path integral directly and does not suffer from difficulties and ambiguities associated with complex contours of Euclidean path integrals for quantum gravity. In addition if fermions are present  quantum computing can evade the sign problem by representing fermions directly on the quantum computer which is another affliction that can cause difficulty for classical computing. Also the representation of black hole states in terms of qubits may yield insight into black hole and de Sitter entropy and how they emerge as a counting of microstates.  Thus it is important  to investigate the use of quantum computing to simulate both black holes and quantum cosmology on a quantum computer.

This paper is organized as follows. In section two we review some aspects of the  Schwarzschild-de Sitter spacetime which highlights many of the conceptual problems inherent for black holes in a Universe with a positive cosmological constant. In section three we review the Kantowski-Sachs cosmology and the special nature of the Narai spacetime. In section four we review quantum computing and how it can be used to simulate quantum mechanical models. In section five we describe the Hamiltonians and Mass operators  of the Black hole interior and Kantowski-Sach cosmologies and how these can be represented in terms of qubit operators on a quantum computer. We show how one can apply the Variational Quantum Eigensolver (VQE) hybrid  quantum-classical algorithms to these systems. We give the results of our calculations for 4,6 and 8 qubits quantum computations. In section 6 we discuss the accuracy of our results, our conclusions and directions for further research.

\section{Schwarzschild-de Sitter Black Hole}

Although the black hole temperature is too small to be measured and is smaller than the cosmological background temperature, the black hole entropy of black holes is quite large and can be tested using black hole mergers \cite{Isi:2020tac}. The cosmological horizon entropy is even larger and is proportional to the inverse of the small ( but observable ) cosmological constant. These large amounts of entropy in the Universe are quite mysterious and need further investigation to understand their origins quantum mechanically.  The Schwarzschild-de Sitter solution has both a cosmological and black hole horizon and is thus conceptually very interesting. The metric of the Schwarschild-de Sitter space-time is given by \cite{Shankaranarayanan:2003ya}:
\begin{equation}d{s^2} =  - \left( {1 - \frac{{2M}}{r} - \frac{{{r^2}}}{{{\ell ^2}}}} \right)d{t^2} + {\left( {1 - \frac{{2M}}{r} - \frac{{{r^2}}}{{{\ell ^2}}}} \right)^{ - 1}}d{r^2} + {r^2}d{\Omega ^2}\end{equation}
\begin{figure}
\centering
  \includegraphics[width = .5 \linewidth]{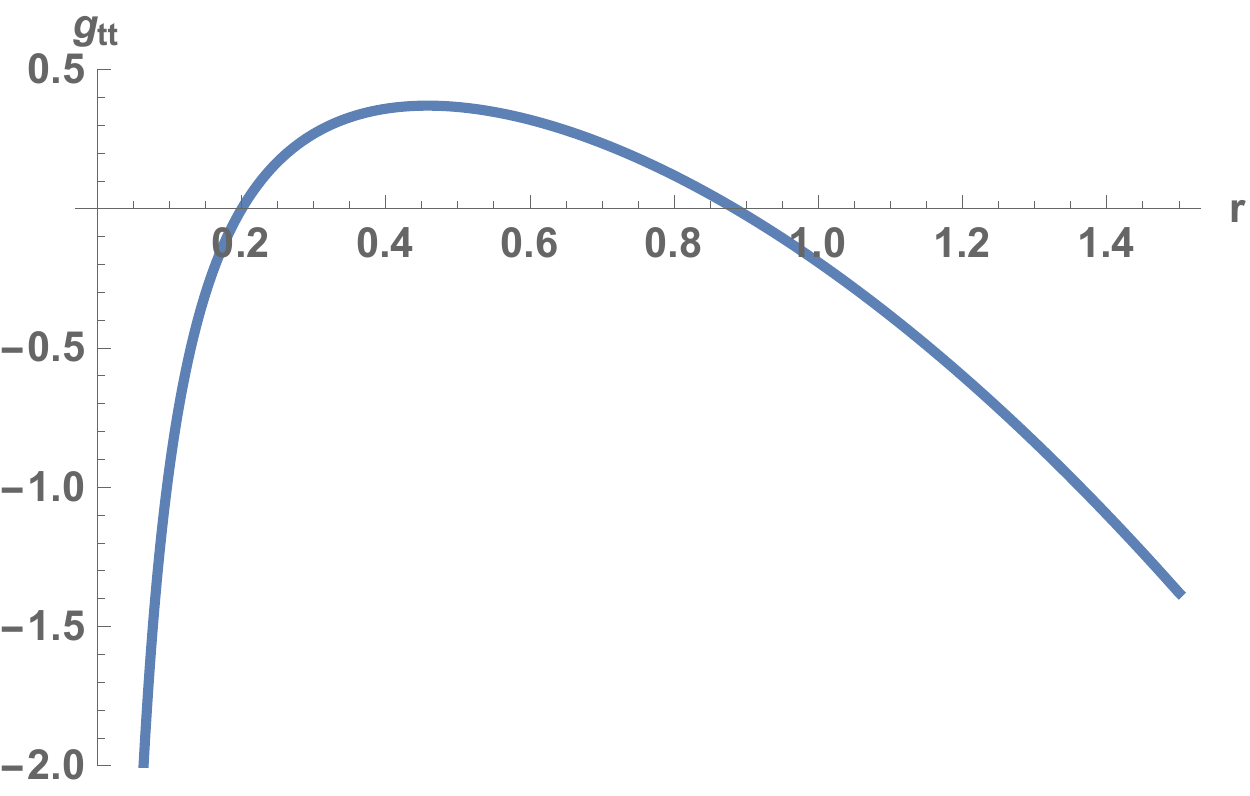}
  \caption{$g_{tt}$ component of the Schwarzschild-de Sitter black hole. The metric component intercepts zero at the black hole and cosmological horizon.
  }
  \label{fig:Radion Potential}
\end{figure}
where $\ell^{-2} = \frac{\lambda}{3}$ and $\lambda$ is the positive cosmological constant. 
The horizons are determined by the positive values of $r$ such that $g_{tt}=0$. These are plotted in figure 1 and are given by:
\[{r_{bh}} = \frac{{2M}}{{\sqrt {3\frac{{{M^2}}}{{{\ell ^2}}}} }}{\mathop{\rm Cos}\nolimits} \left[ {\frac{1}{3}\left( {\pi  + {\mathop{\rm Arccos}\nolimits} \left[ {3\sqrt {3\frac{{{M^2}}}{{{\ell ^2}}}} } \right]} \right)} \right]\]
\begin{equation}{r_{ch}} = \frac{{2M}}{{\sqrt {3\frac{{{M^2}}}{{{\ell ^2}}}} }}{\mathop{\rm Cos}\nolimits} \left[ {\frac{1}{3}\left( {\pi  - {\mathop{\rm Arccos}\nolimits} \left[ {3\sqrt {3\frac{{{M^2}}}{{{\ell ^2}}}} } \right]} \right)} \right]\end{equation}
where $r_{bh}$ is the black hole horizon and $r_{ch}$ is the cosmological horizon. The entropy of these horizons are given by one quarter the horizon area and are:
\[{S_{bh}} = \pi {\left( {\frac{{2M}}{{\sqrt {3\frac{{{M^2}}}{{{\ell ^2}}}} }}{\mathop{\rm Cos}\nolimits} \left[ {\frac{1}{3}\left( {\pi  + {\mathop{\rm Arccos}\nolimits} \left[ {3\sqrt {3\frac{{{M^2}}}{{{\ell ^2}}}} } \right]} \right)} \right]} \right)^2}\]
\begin{equation}{S_{ch}} = \pi {\left( {\frac{{2M}}{{\sqrt {3\frac{{{M^2}}}{{{\ell ^2}}}} }}{\mathop{\rm Cos}\nolimits} \left[ {\frac{1}{3}\left( {\pi  - {\mathop{\rm Arccos}\nolimits} \left[ {3\sqrt {3\frac{{{M^2}}}{{{\ell ^2}}}} } \right]} \right)} \right]} \right)^2}\end{equation}
These are plotted in figure 2. The sum of the cosmological and black hole horizon entropies is given by:
\[{S_{tot}} = {S_{bh}} + {S_{ch}}\]
and is plotted in figure 3.

The inverse temperature associated with these horizons can be determined from
\begin{equation}\beta  = \frac{{\partial S}}{{\partial M}}\end{equation}
and is plotted in figure 4. The temperature is plotted in figure 5. For samll values of $M \ll \ell$ the entropy and temperature have the expansions
\begin{align}
&{S_{bh}} = 4\pi {M^2} + \frac{{32\pi {M^4}}}{{{\ell ^2}}} +  \ldots \nonumber\\
&{S_{ch}} = \pi {\ell ^2} - 2\pi M\ell  +  \ldots \nonumber\\
&{T_{bh}} = \frac{1}{{8\pi M}} - \frac{{2M}}{\pi \ell^2} +  \ldots \nonumber \\
&{T_{ch}} =  - \frac{1}{{2\pi \ell }} + \frac{M}{\pi \ell^2} +  \ldots \nonumber\\
&\beta_{bh} = 8 \pi  M + \frac{128 \pi M^3}{\ell^2} + \ldots \nonumber\\
&\beta_{ch} = -2\pi \ell - 4 \pi M + \ldots
\end{align}
For Nariai spacetime the two horizons overlap. Both horizons are located at $r_N= \frac{1}{\sqrt{\lambda}}$. It has zero temperture and entropy $2/3$ that of de Sitter space or $S_N = \frac{2\pi}{\lambda}$. The mass of the Nariai spacetime is given by  $M_N= \frac{\ell}{\sqrt{27}} = \frac{1}{3\sqrt{\lambda}}$.

Another interesting feature of the Schwarzschild-de Sitter space-time is that because of the maximum value for the mass of the black hole the partition function defined by
\begin{equation}Z(\beta ) =  \int_0^{\sqrt {1/27} } {dM{e^{S(M)}}} {e^{ - \beta M}} =\int_0^{\sqrt {1/27} } {dM{e^{\pi {{\left( {\frac{2}{{\sqrt 3 }}{\mathop{\rm Cos}\nolimits} [\frac{1}{3}\left( {\pi  + {\mathop{\rm Arccos}\nolimits} \left( {\sqrt {27} M} \right)} \right)]} \right)}^2}}}} {e^{ - \beta M}}\end{equation}
converges for all values of $\beta$, unlike the Schwarzschild case, where it is defined by a saddle point \cite{Ginsparg:1982rs}
\cite{Gross:1982cv}. For simplicity in the above we have set $\lambda =3$ so that $\ell = 1$. Various thermodynamic quantities for the different spacetimes are listed in table 1. 

\begin{table}[h]
\centering
\begin{tabular}{|l|l|l|l|}
\hline
Spacetime       & Entropy  & T & $\beta$\\ \hline
de Sitter   &   $\frac{3\pi}{\lambda}$ &  $ \frac{1}{{2\pi }}\sqrt {\frac{\lambda }{3}} $&  $2\pi\sqrt{\frac{3}{\lambda}}$      \\ \hline
Nariai & $\frac{2\pi}{\lambda}$ &  $0$ & $\infty$ \\ \hline
Schwarzschild   &   $4 \pi M^2$  &  $ \frac{1}{8\pi M} $ & $8 \pi M $\\ \hline
Schwarzschild-de Sitter    &  $ 4\pi {M^2} + \frac{{32\pi \lambda {M^4}}}{{{3}}} +  \ldots$ & $\frac{1}{{8\pi M}} - \frac{{2\lambda M }}{3\pi } +  \ldots $ & $8\pi M + \frac{{128\pi \lambda {M^3}}}{3} +  \ldots $   \\ \hline

\end{tabular}
\caption{ Entropy, temperature and inverse temperature for de Sitter, Nariai, Schwarzschild and Schwarzschild-de Stter spacetimes with black hole mass $M$ and cosmological constant $\lambda$.}
\end{table}

\begin{figure}
\centering
  \includegraphics[width = .5 \linewidth]{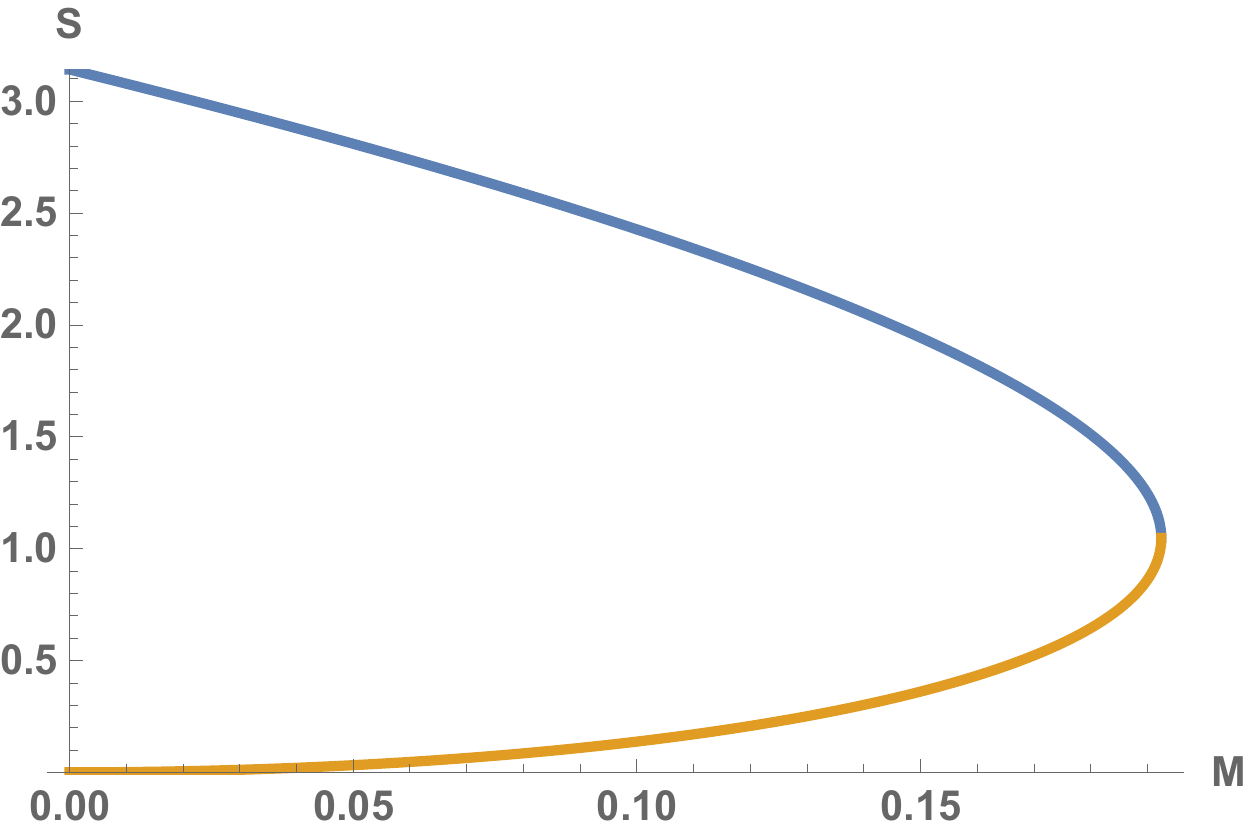}
  \caption{Entropy versus Mass for Schwarzschild-de Sitter spacetime. Blue curve indicates the entropy from the cosmological horizon and orange curve indicates the entropy of the black hole horizon.
  }
  \label{fig:Radion Potential}
\end{figure}

\begin{figure}
\centering
  \includegraphics[width = .5 \linewidth]{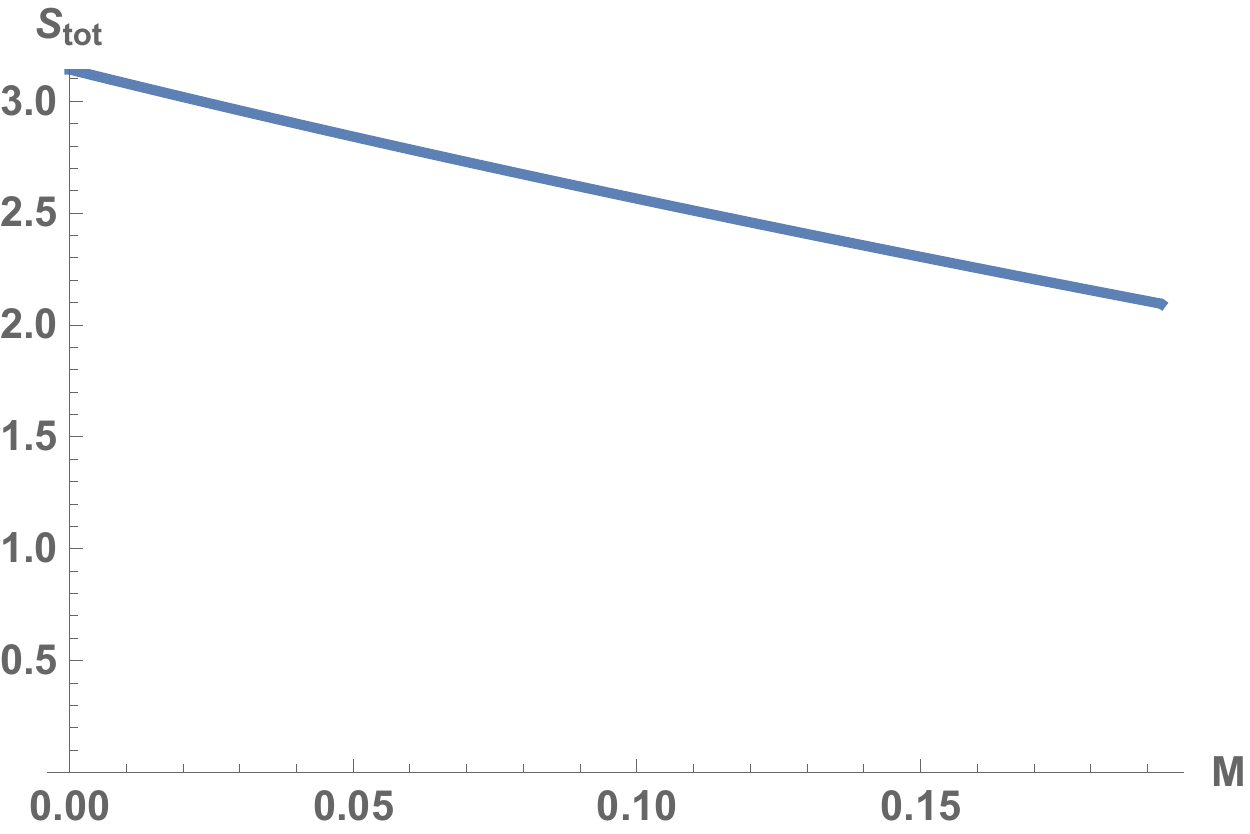}
  \caption{Total Entropy versus Mass for Schwarzschild-de Sitter spacetime.
  }
  \label{fig:Radion Potential}
\end{figure}

\begin{figure}
\centering
  \includegraphics[width = .5 \linewidth]{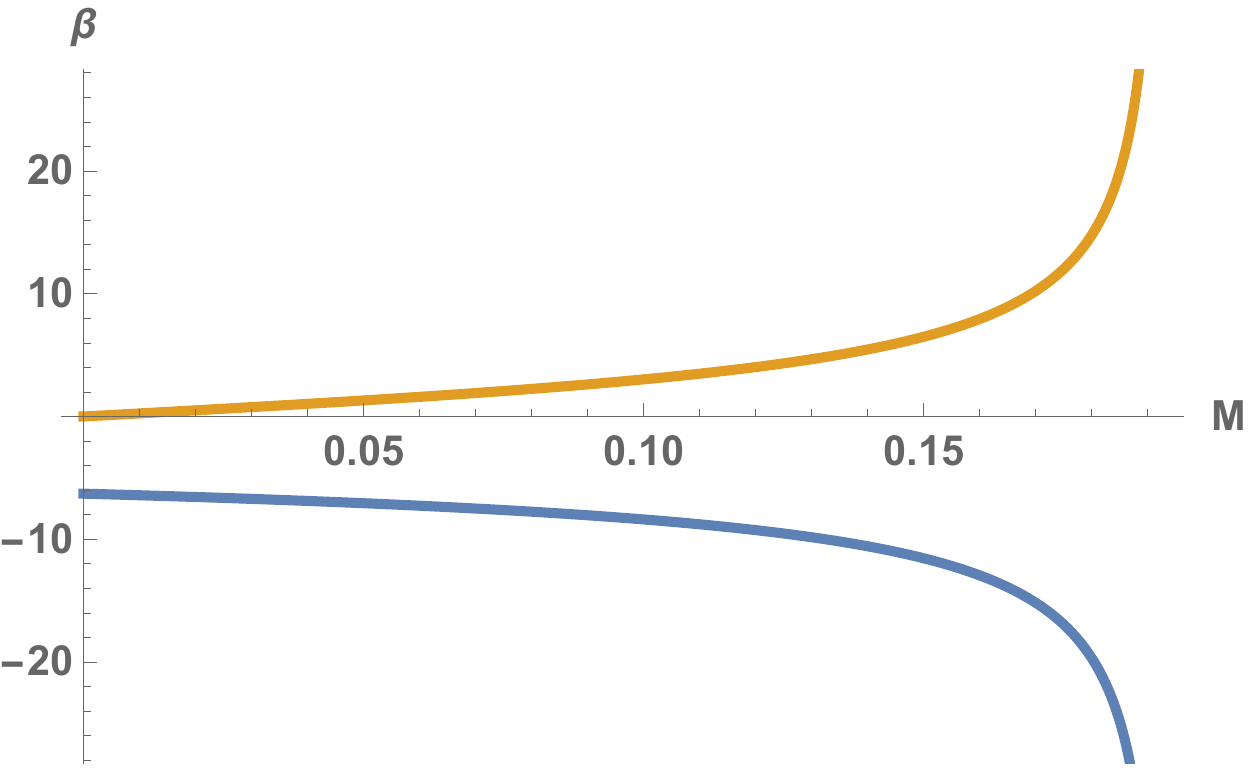}
  \caption{Inverse temperature versus Mass for Schwarzschild-de Sitter spacetime. Blue curve indicates the cosmological horizon and orange curve indicates the black hole horizon.
  }
  \label{fig:Radion Potential}
\end{figure}

\begin{figure}
\centering
  \includegraphics[width = .5 \linewidth]{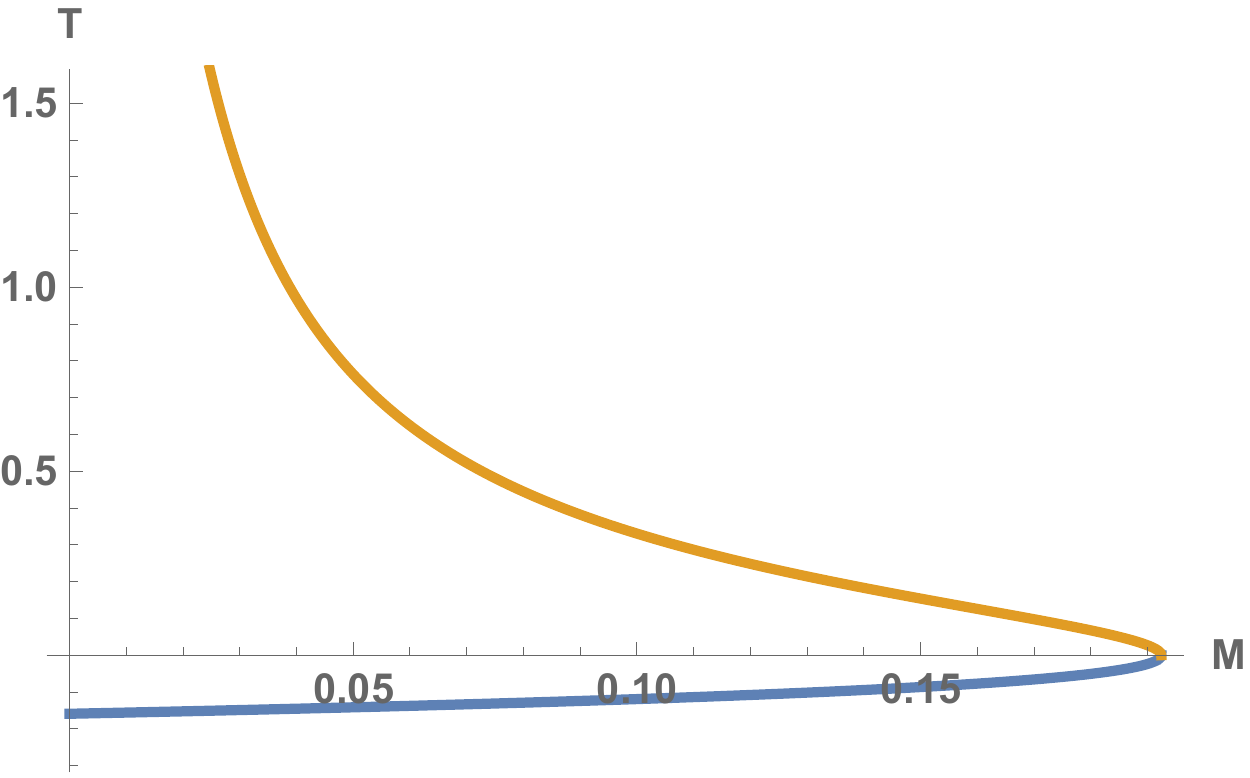}
  \caption{Temperature versus Mass for Schwarzschild-de Sitter spacetime. Blue curve indicates the cosmological horizon and orange curve indicates the black hole horizon.
  }
  \label{fig:Radion Potential}
\end{figure}

\newpage


\section{ Hamiltonian and Mass Operators of \\
Schwarzschild-de Sitter and Kantowski-Sachs cosmology}

The Kantowski-Sachs cosmology is an anistropic spacetime with spatial topology $S^1 \times S^2$. The main operators we will need for both the the Schwarszschild-de Sitter and the Kantowski-Sachs cosmology in this paper are the Hamiltonian and Mass operators.
Using the metric ansatz for the spherical model \cite{Berger:1972pg}
\cite{Kenmoku:1998he}
\cite{Fischler:1990pk}
\cite{Kiefer:1998mi} and following Fischler, Morgan and Polchinski \cite{Fischler:1990pk} this is given by:
\begin{equation}d{s^2} =  - {N^2}d{t^2} + {a(r,t)^2}d{r^2} + {b(r,t)^2}d{\Omega_2^2}\end{equation}
The Hamiltonian, momentum and mass constraints for the spherical model are
\begin{align}
&H = -\frac{{Gap_a^2}}{{2{b^2}}} + \frac{{G{p_a}{p_b}}}{b} - \frac{1}{{2G}}\left\{ {\frac{{b'b'}}{a} - \frac{{2b}}{{{a^2}}}a'b' + \frac{{2b}}{a}b'' - a + \lambda a{b^2}} \right\} \nonumber\\
&P =  - a'{p_a} + b'{p_b} \nonumber\\
&M = \frac{{Gp_a^2}}{{2b}} + \frac{b}{{2G}}\left\{ {1 - \frac{{b'b'}}{{{a^2}}} - \frac{{\lambda {b^2}}}{3}} \right\}
\end{align}
The operators are essentially identical for the Kantowski-Sachs cosmology and black hole interior. The main difference is that in the Kantowski-Sachs cosmology the variable $r$ parametrizes a circle whereas for the black hole interior it parametrizes an interval.

\subsection*{$a,b$ representation}
 The $a,b$ representation gives a clear physical picture of the metric components in terms of radii. We will leave the full inhomogeneous midisupersuspace for future investigations and consider the simpler minisuperspace. Using the metric ansatz: 
\begin{equation}d{s^2} =  - {N^2}d{t^2} + {a(t)^2}d{r^2} + {b(t)^2}d{\Omega_2 ^2}\end{equation}
the Hamiltonian constraint and mass operator in the $a,b$ representation are give by:
\begin{align}
&H =   -\frac{a}{{{2b^2}}}p_a^2 + \frac{1}{b}{p_a}{p_b} + \frac{a}{2} - \frac{\lambda}{2} a{b^2}\nonumber\\
&M = \frac{1}{2b}p_a^2 + \frac{b}{2} - \frac{\lambda }{6}b^3
\end{align}
Where we have set $G=1$.Using these definitions the Hamiltonian constraint and mass operator obey the relation:
\begin{equation}[{H},M] = 0\end{equation}
For the Kantowski-Sachs cosmology one has spatial topology $S^1 \times S^2$ and the metric ansatz:
\begin{equation}d{s^2} =  - {N^2}d{t^2} + {a^2}(t)d\Omega _1^2 + {b^2}(t)d\Omega _2^2\end{equation}
 One has a curvature term, but only for the $S^2$ of radius $b$. The Lagrangian, Hamiltonian, Mass and canonical momentum are given by \cite{Conradi:1994yy}:
 \begin{align}
&L =  - \frac{{b\dot b\dot a}}{N} - \frac{{a{{\dot b}^2}}}{{2N}} + \frac{{Na}}{2} - {\frac{{\lambda Nab}}{2}^2} \nonumber\\
& H = \frac{{a{{\dot b}^2}}}{{{N^2}}} + 2\frac{{b\dot a\dot b}}{{{N^2}}} + \frac{{2 \cdot 1}}{2}b - a{b^2}\lambda  = 0 \nonumber\\
&M = \frac{{b{{\dot b}^2}}}{{2{N^2}}} + \frac{b}{2} - \frac{\lambda }{6}{b^3}\nonumber \\
&{p_a} =  - \frac{{b\dot b}}{N}\nonumber \\
&{p_b} =  - \frac{{b\dot a}}{N} - \frac{{a\dot b}}{N}
\end{align}

\subsection*{$u,v$ representation}
Although the $a,b$ representation have a clear physical interpretation it turns out the $u,v$ representation \cite{Jalalzadeh:2011yp}
\cite{Djordjevic:2015uga} \cite{Ali:2018vmt}
 \cite{Kiefer:2009zwx} is more convenient for quantum computing
%
Defining
\[u = {b^{1/2}}(a + 1)\]
\begin{equation}v = {b^{1/2}}(a - 1)\end{equation}
so that
\begin{align}
&a = \frac{{u + v}}{{u - v}}\nonumber\\
&b = \frac{1}{4}{(u - v)^2}    
\end{align}
the Hamiltonian constraint $H$ and Mass operator $M$ in the $u,v$ variables are
\begin{align}
2b H &= \frac{1}{2}\left( {p_u^2 - p_v^2} \right) + \frac{1}{2}\left( {{u^2} - {v^2}} \right) - \frac{\lambda }{32}\left( {{u^2} - {v^2}} \right){\left( {u - v} \right)^4} \nonumber\\
4M &= \frac{1}{2}{\left( {{p_u} + {p_v}} \right)^2} + \frac{1}{2}{\left( {u - v} \right)^2} - \frac{\lambda}{96}{\left( {u - v} \right)^6}
\end{align}
In the above the $H$ of the $a,b$ representation  was scaled by $2b$ to obtain the $H$ of the $u,v$ representation. The $M$ of the $a,b$ representation was scaled by $4$ to obtain the $M$ of the $u,v$ representation.  This form is useful when one considers the quantum computation of the Schwarzschild-de Sitter model. Physical states satisfy:
\begin{align}
&{2b H}\left| \psi  \right\rangle  = 0 \nonumber\\
&4M\left| \psi  \right\rangle = 4m_{bh} \left| \psi  \right\rangle 
\end{align}
with $m_{bh}$ the black hole mass. These are the equations we will solve using quantum computing.

\subsection*{Application of Mass operator to Nariai spacetime}

One can apply the mass operator to the Nariai spacetime. Consider the special solution to the Kantowki-Sachs cosmology with metric ansatz:
\begin{equation}d{s^2} =  - {N^2}d{t^2} + {a^2}d{z^2} + {b^2}d{\Omega ^2}\end{equation}
given by:
\begin{equation}d{s^2} =  - d{t^2} + {e^{2t\sqrt \lambda  }}d{z^2} + \frac{1}{\lambda }d{\Omega ^2}\end{equation}
so that
\begin{equation}N(t) = 1,a(t) = {e^{t\sqrt \lambda  }},b(t) = \frac{1}{{\sqrt \lambda  }}\end{equation}
The mass operator is given by:
\begin{equation}2M = \frac{{p_a^2}}{b} + b - \frac{\lambda }{3}{b^3}\end{equation}
with $p_a$ the canonical momentum associated with $a$ given by:
\begin{align}
& {p_a} =  - \frac{{b\dot b}}{N} \nonumber\\
& p_b = - \frac{a \dot b}{N} + \frac{b \dot a }{N}
\end{align}
Using the above solution we have
$p_a=0$
so the mass operator becomes:
\begin{equation}2M = 0 + \frac{1}{{\sqrt \lambda  }} - \frac{\lambda }{3}\frac{1}{{{\lambda ^{2/3}}}}\end{equation}
so that the Narain mass is given by:
\begin{equation}M_N = \frac{1}{3}\frac{1}{{\sqrt \lambda  }}\end{equation}
using the definition 
\begin{equation}\lambda  = 3{\ell ^{ - 2}}\end{equation}
we have:
\begin{equation}M_N = \frac{1}{{\sqrt {27} }}\ell \end{equation}
which corresponds to the mass parameter for the Nariai space time. It is clear from (3.13) that the Nariai space time has no singularity being equivalent to $dS_2\times S^2$.  The Nariai space time has topology $R\times S^1 \times S^2$ which is distinct from the topology of de Sitter which is $R\times S^3$. The topology of these spaces is different although connected through the nontrivial topological Hopf mapping relating the fibration of $S^1\times S^2$ to $S^3$.  It is also interesting to consider magnetically charged Kantowski-Sachs spacetimes which may be more stable and long lived as the radius $b$ is potentially stabilized by the magnetic flux \cite{Das:2014ria}
\cite{Katore:2012upa}
\cite{Katore:2010zz}
\cite{Fornal:2011tw}
\cite{Arnold:2010qz}
\cite{Arkani-Hamed:2007ryu}
\cite{Bousso:1996pn}. 

The instability of the Nariai spacetime can also be seen form the potential plot in figure 6. The Schwarzschild potential is defined from:
\begin{equation}4M = \frac{1}{2}{\left( {{p_u} + {p_v}} \right)^2} + {V_S}(u - v)\end{equation}
with
\begin{equation}{V_S}(u - v) = \frac{1}{2}{\left( {u - v} \right)^2}\end{equation}
which is plotted in figure 6 as the upward curving parabolic curve. The Schwarzschild-de Sitter potential is defined from:
\begin{equation}4M = \frac{1}{2}{\left( {{p_u} + {p_v}} \right)^2} + {V_{SD}}(u - v)\end{equation}
with 
\begin{equation}{V_{SD}}(u - v) = \frac{1}{2}{\left( {u - v} \right)^2} - \frac{\lambda }{{96}}{(u - v)^2}\end{equation}
which plotted as the blue curve in figure 6 which curves downward for large $b$ or $u-v$. The maximum of this potential is the Nariai point and is unstable to $b$ or $u-v$ contracting to zero or expanding to infinity on the left or right respectively of the maximum of the potential. 

The same behavior of the Nariai point can be seen in contour plots. In the $a,b$ representation the contour in $p_a,b$ for:
\begin{equation}M = \frac{{p_a^2}}{{2b}} + \frac{b}{2} - \frac{{\lambda {b^3}}}{6}\end{equation}
with $M < M_N$ to the left and $M=M_N$ to the right in figure 7. The Narai point which is at intersection of the loop part of the contour with the expanding part. For masses less that the Nariai mass there is a barrier which separates these two parts of the contour but for the Nariai spacetiem there is no barrier and the Nariai spacetime sits at the cross in a precarious position between contraction and expansion. The same behaviour can be seen in the $u,v$ representation where we plot the contour:
\begin{equation}4M = \frac{1}{2}{\left( {{p_u} + {p_v}} \right)^2} + \frac{1}{2}{\left( {u - v} \right)^2} - \frac{\lambda }{{96}}{(u - v)^6}\end{equation}
in figure 8 for $M < M_N$ to the left and $M=M_N$ to the right. The cross between the contracting and expanding $u-v$ gives the unstable Narai point.

\begin{figure}
\centering
  \includegraphics[width = .5 \linewidth]{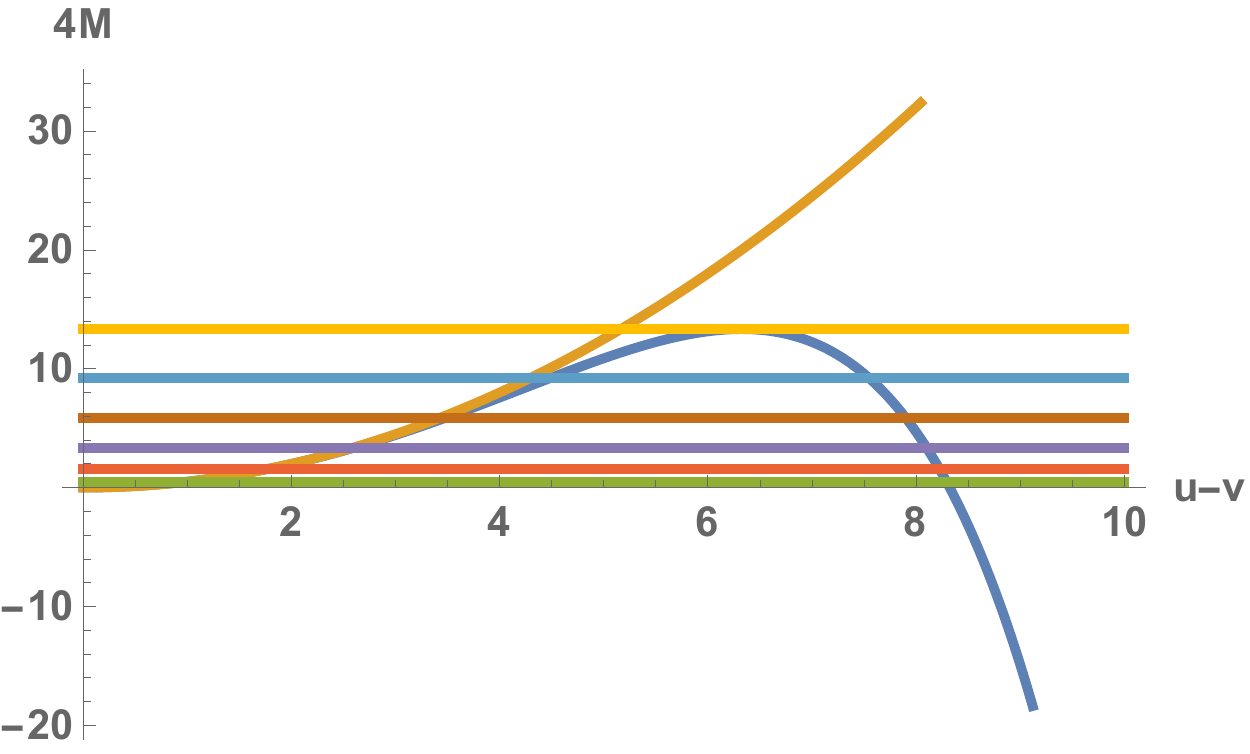}
  \caption{The upward curving parabolic curve indicated the Schwarzschild potential in the $u,v$ representation. The blue curve which slopes down at large $u-v$ represents the Schwarzschild-de Sitter solution which we plot for $\lambda=.01$. The point on top of the potential is the Nariai point which appears to be unstable. Various eigenvalues of the mass operator are plotted as horizontal lines with the Nariai mass intersecting the top of the potential denoted with the yellow line. 
  }
  \label{fig:Radion Potential}
\end{figure}

\begin{figure}[!htb]
\centering
\minipage{0.5\textwidth}
  \includegraphics[width=\linewidth]{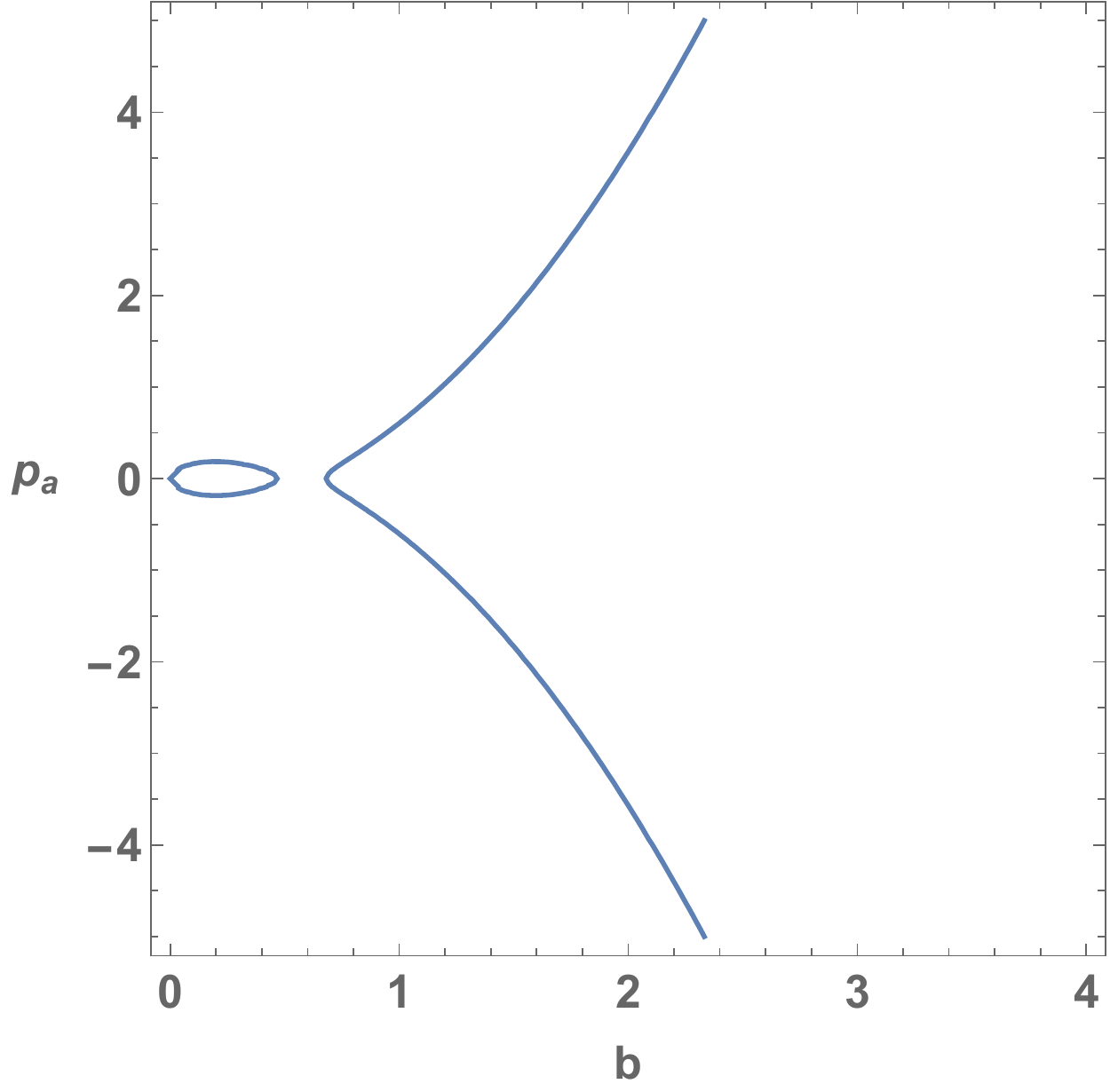}
\endminipage\hfill
\minipage{0.5\textwidth}
  \includegraphics[width=\linewidth]{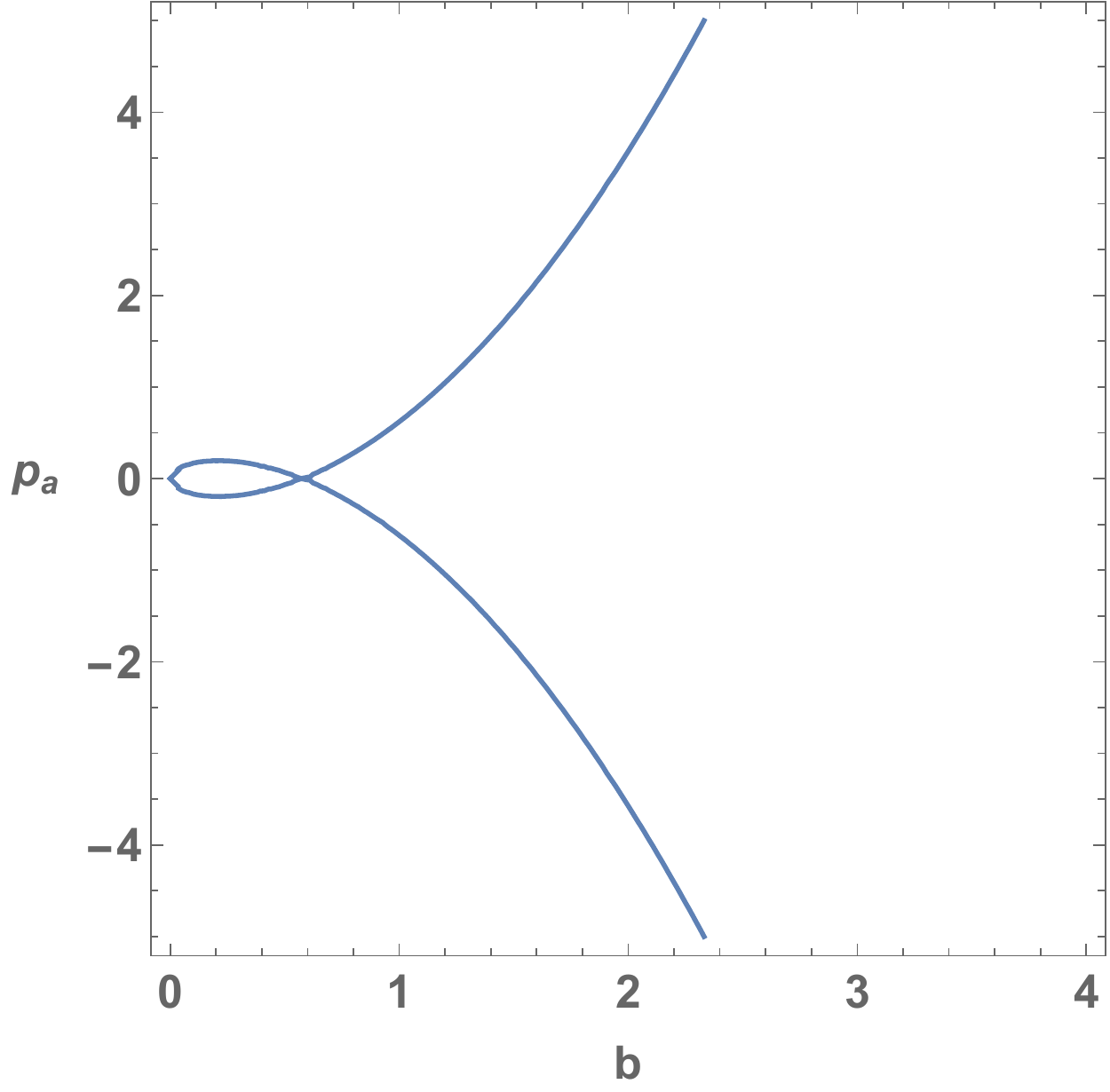}
\endminipage\hfill
\caption{(Left) Contour plot in the $a,b$ representation of $M = \frac{{p_a^2}}{{2b}} + \frac{b}{2} - \frac{{\lambda {b^3}}}{6}$ for a mass (left) less than the Narai mass (right) equal to the Narai mass . Note that at the Narai point where the curves cross there is no barrier separating contracting and expanding curves in $b$.}
\end{figure}

\begin{figure}[!htb]
\centering
\minipage{0.5\textwidth}
  \includegraphics[width=\linewidth]{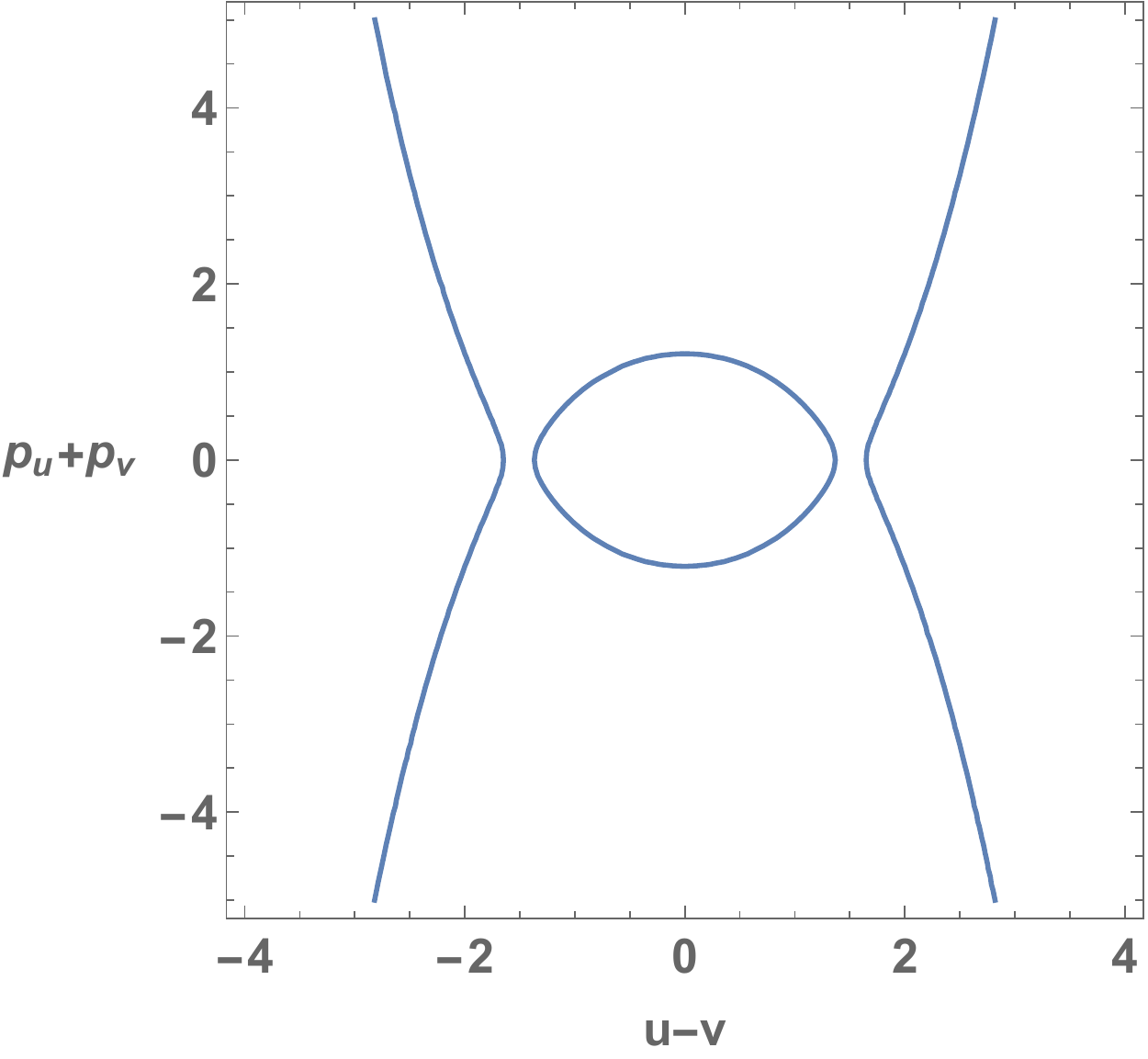}
\endminipage\hfill
\minipage{0.5\textwidth}
  \includegraphics[width=\linewidth]{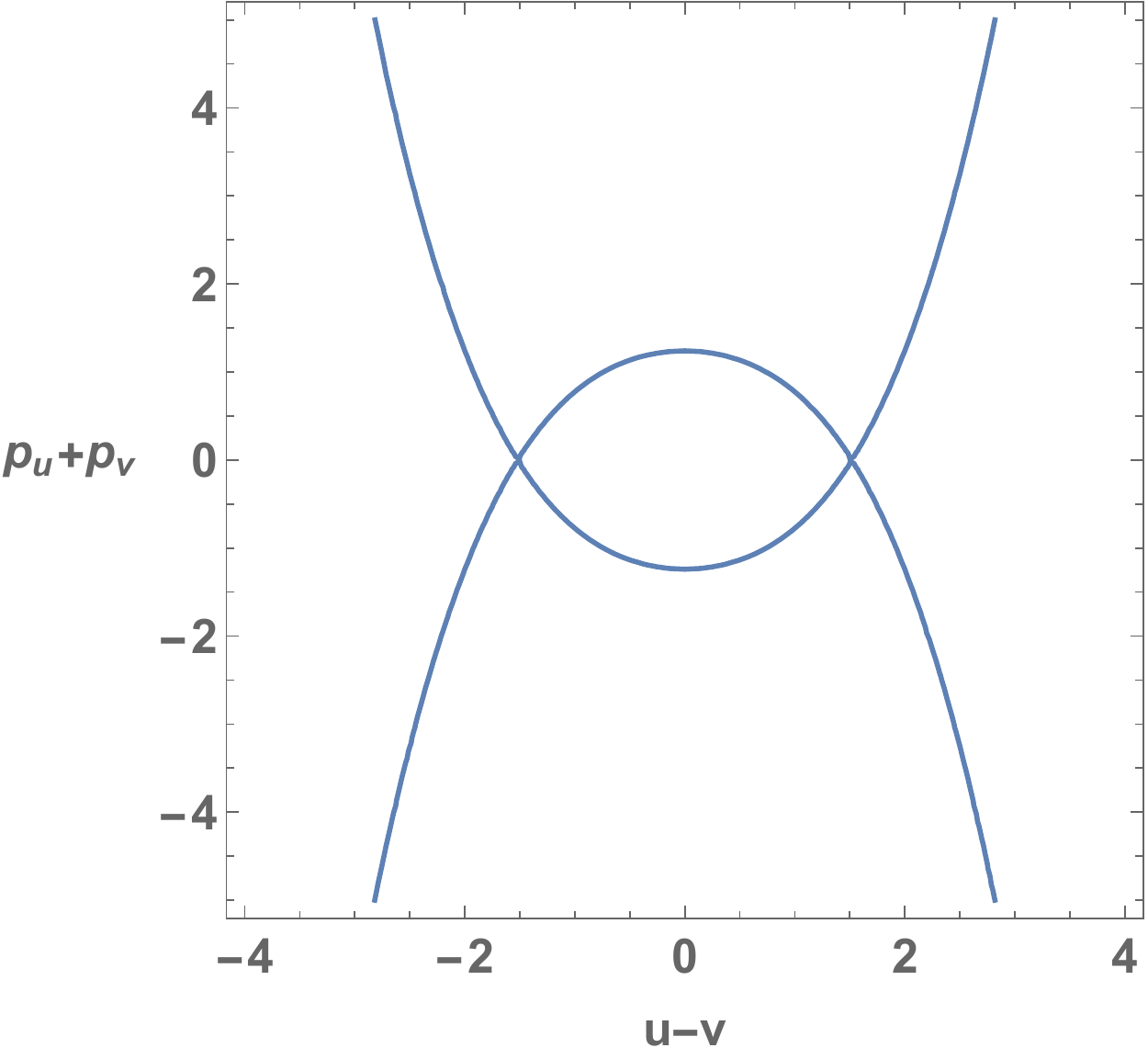}
\endminipage\hfill
\caption{Contour plot in the $u,v$ representation of $4M = \frac{1}{2}{\left( {{p_u} + {p_v}} \right)^2} + \frac{1}{2}{\left( {u - v} \right)^2} - \frac{\lambda }{{96}}{(u - v)^6}$ for a mass (left) less than the Narai mass (right) equal to the Narai mass . Note that at the Narai point where the curves cross there is no barrier separating contracting and expanding curves in $u-v$.}
\end{figure}

\section{Quantum Computing}

Quantum computing can be used to simulate quantum systems. In particular it can be used to simulate quantum cosmology or black holes. In quantum computing one represents the Hilbert space of quantum states in terms of qubits. One represents the Hamiltonian, canonical momentum and other operators in terms of a basis which maps the Hamiltonian to an operator on qubits through a process called mapping. The four types of basis we consider are the position basis, the oscillator basis, the finite difference basis and creation-annihilation operator basis which we will describe. After one maps the Hamiltonian to qubits one expands the Hamiltonian in terms of tensor products of $2\times 2$ Pauli matrices (denoted $X,Y,Z,I$ ) or ($\sigma_x,\sigma_y,\sigma_z,\sigma_0$). These tensor products are called Pauli terms and can be represented efficiently on a quantum computer in terms of gates. Finally one chooses a quantum algorithm to run on the quantum computer to simulate the physical system. In this paper we consider the Variational Quantum Eigensolver (VQE) which is hybrid classical-quantum algorithm that determines a variational estimate to the ground state energy and ground state wave function of the system. The VQE works by combining a classical optimization of parameters for associated gates describing the ground state wave function with expectation values of the Hamiltonian being done on the quantum computer and optimizations evaluated on a classical computer  in a loop until convergence is achieved. 

\subsection*{Gaussian or Simple Harmonic Oscillator basis}

The first step in quantum computing simulation is to choose a representation or basis for the Hamiltonian. 
The Gaussian or Harmonic Oscillator representation is a very useful basis based on the matrix treatment of the simple harmonic oscillator which is sparse in representing both the position and momentum operators. For the position operator we have:
\begin{equation} 
 Q_{osc} = \frac{1}{\sqrt{2}}\begin{bmatrix}
 
   0 & {\sqrt 1 } & 0 &  \cdots  & 0  \\ 
   {\sqrt 1 } & 0 & {\sqrt 2 } &  \cdots  & 0  \\ 
   0 & {\sqrt 2 } &  \ddots  &  \ddots  & 0  \\ 
   0 & 0 &  \ddots  & 0 & {\sqrt {N-1} }  \\ 
   0 & 0 &  \cdots  & {\sqrt {N-1} } & 0  \\ 
\end{bmatrix}
  \end{equation}
while for the momentum operator we have:
\begin{equation}
 P_{osc} = \frac{i}{\sqrt{2}}\begin{bmatrix}
 
   0 & -{\sqrt 1 } & 0 &  \cdots  & 0  \\ 
   {\sqrt 1 } & 0 & -{\sqrt 2 } &  \cdots  & 0  \\ 
   0 & {\sqrt 2 } &  \ddots  &  \ddots  & 0  \\ 
   0 & 0 &  \ddots  & 0 & -{\sqrt {N-1} }  \\ 
   0 & 0 &  \cdots  & {\sqrt {N-1} } & 0  \\ 
\end{bmatrix}
  \end{equation}
The Hamiltonian is then represented as $H(q,p) = H( Q_{osc}, P_{osc})$ and is expanded interms of qubits.

\subsection*{Position basis}

In the position basis the position matrix is diagonal but the momentum matrix is dense and constructed from the position operator using a Sylvester matrix $F$. In the position basis the position matrix is:
\begin{equation}
{\left( {{Q_{pos}}} \right)_{j,k}} = \sqrt {\frac{{2\pi }}{{4N}}} (2j - (N + 1)){\delta _{j,k}}
\end{equation}
and the momentum matrix is:
\begin{equation}{P_{pos}} = {F^\dag }{Q_{pos}}F\end{equation}
where 
\begin{equation}{F_{j,k}} = \frac{1}{{\sqrt N }}{e^{\frac{{2\pi i}}{{4N}}(2j - (N + 1))(2k - (N + 1))}}\end{equation}
The Hamiltonian is then represented as $H(q,p) = H( Q_{pos}, P_{pos})$

\subsection*{Finite difference basis}

 In the finite difference basis the position matrix is:
\begin{equation}{\left( {{Q_{fd}}} \right)_{j,k}} = \sqrt {\frac{1}{{2N}}} (2j - (N + 1)){\delta _{j,k}}\end{equation}
and the momentum-squared matrix is:
\begin{equation} 
 P_{fd}^2 = \frac{N}{2}\begin{bmatrix}
 
   2 & - 1  & 0 &  \cdots  & 0  \\ 
   -1 & 2 & -1 &  \cdots  & 0  \\ 
   0 & -1 &  \ddots  &  \ddots  & 0  \\ 
   0 & 0 &  \ddots  & 2 & -1  \\ 
   0 & 0 &  \cdots  & -1 & 2  \\ 
\end{bmatrix}
  \end{equation}
The Hamiltonian is then represented as $H(q,p) = H( Q_{fd}, P_{fd})$

\subsection*{Creation and Annihilation  operator basis}

The creation and annihilation operator basis is similar to the oscillator basis. One writes
\begin{align}
&A = \frac{1}{{\sqrt 2 }}(i P_{osc} + Q_{osc}) \nonumber \\
&{A^\dag } = \frac{1}{{\sqrt 2 }}( - i P_{osc} + Q_{osc} )
\end{align}
Then the Hamiltonian is simply:
\begin{equation}{H_(a, a^\dag) = H(A,  A^\dag})\end{equation}
where here $a$ and $a^\dag$ are creationa and annihilation operators.

\subsection*{Multiple variables}
In this paper we will use the oscillator basis and tensor products for two variables which we denote as $u,v$. The can be constructed through tensor products as 
\[u = {Q_{osc}} \otimes {I_{{2^{\frac{q}{2}}}}}\]
\[v = {I_{_{{2^{\frac{q}{2}}}}}} \otimes {Q_{osc}}\]
\[{p_u} = {P_{osc}} \otimes {I_{{2^{\frac{q}{2}}}}}\]
\begin{equation}{p_v} = {I_{_{{2^{\frac{q}{2}}}}}} \otimes {P_{osc}}\end{equation}
where $q$ is the number of qubits so the matrices will be of size $2^q \times 2^q$. The Hamiltonian or Mass operator is then a function of these variables and is expanded in terms of Pauli terms as:
\begin{equation}H(u,v,{p_u},{p_v}) = \sum\limits_n {{a_n}{P_n}} \end{equation}
where 
\begin{equation}{P_n} = {\sigma _i} \otimes {\sigma _j} \otimes  \ldots {\sigma _k}\end{equation}
and $i,j,k,\ldots$ go from 0 to 3.

\section{Quantum Computing Results}

Although there are many quantum algorithms available we concentrate of the Variational Quantuntum Eigensolver (VQE) \cite{Tilly:2021jem}\cite{vqe} as it has been shown that this algorithm can be implemented on near term quantum computers and can be made resistant to noise through noise mitigation methods. The VQE algorithm calculates the lowest eigenvalue and ground state of an operator by using the variational method of quantum mechanics where the variational wave functions are represented as quantum gates and the variational parameters are rotation angles for these gates. Then these parameters are optimized by continually updating the angles in the gates and evaluating the expectation value of the operator until the process converges. This yields a least upper bound on the lowest eigenvalue which is the final result as indicated in the formula below:
\begin{equation}{m_0} \le \frac{{\left\langle {\psi ({\theta _i}} \right|M\left| {\psi ({\theta _i}} \right\rangle }}{{\left\langle {\psi ({\theta _i}} \right|\left. {\psi ({\theta _i}} \right\rangle }}\end{equation}
where we use the $u,v$ representation of the mass operator:
\begin{equation}4M = \frac{1}{2}{\left( {{p_u} + {p_v}} \right)^2} + \frac{1}{2}{\left( {u - v} \right)^2} - \frac{\lambda}{96}{\left( {u - v} \right)^6}\end{equation}
The optimization of the parameters is done using an optimization algorithm on a classical computer. This is why the VQE is called a hybrid classical-quantum algorithm. An example of a convergence graph for the mass computation using the  Limited-memory BFGS Bound (L-BFGS-B)  optimizer is shown in figure 9 where we see that VQE algorithm quickly converges to the upper bound estimate.  Our results for the VQE computations using the IBM Qiskit state-vector simulator are shown in table 2 and figure 10. The results are very accurate for 4 qubits but the accuracy decreases for 6 and 8 qubits. This indicates that the parametrization of the variational wave forms in terms of quantum gates, for large number of qubits, does not accurately capture the exact mass eigenstates, and one must use a different way to represent and parametrize the states. This may lead to a larger quantum circuit however and a greater circuit depth, to accurately represent the Schwarzschild-de Sitter of Kantowski-Sachs cosmology states.

\begin{table}[ht]
\centering
\begin{tabular}{|l|l|l|l|l|}
\hline
Mass Operator       & No. of Qubits  &  No. Pauli Terms & Exact Result & VQE Result \\ \hline
4M with $\lambda = .01$ & 4 & 57  & 0.0 &  $1.023\times 10^{-10}$  \\ \hline
4M  with $\lambda = .01$ & 6 & 745 & 0.0 & $3.217\times 10^{-1}$     \\ \hline
4M  with $\lambda = .005$ & 8 & 6611 & 0.0 & $ 7.799\times 10^{-1}$    \\ \hline

\end{tabular}
\caption{\label{tab:BasisCompare}  VQE results for the lowest eigenvalue of the Mass operator for Schwarzschild-de Sitter black hole using the oscillator basis. The Hamiltonian was mapped to 4, 6 and 8-qubit operators for $(u,v)$ coordinates in mini-superspace. The quantum circuit for each simulation utilized an \(R_y\) variational form, with a fully entangled circuit of depth 3. The backend used was a state-vector simulator. The VQE results were obtained using the state-vector simulator with no noise and the Limited-memory BFGS Bound (L-BFGS-B)  optimizer.}
\end{table}
\begin{figure}
\centering
  \includegraphics[width = .4 \linewidth]{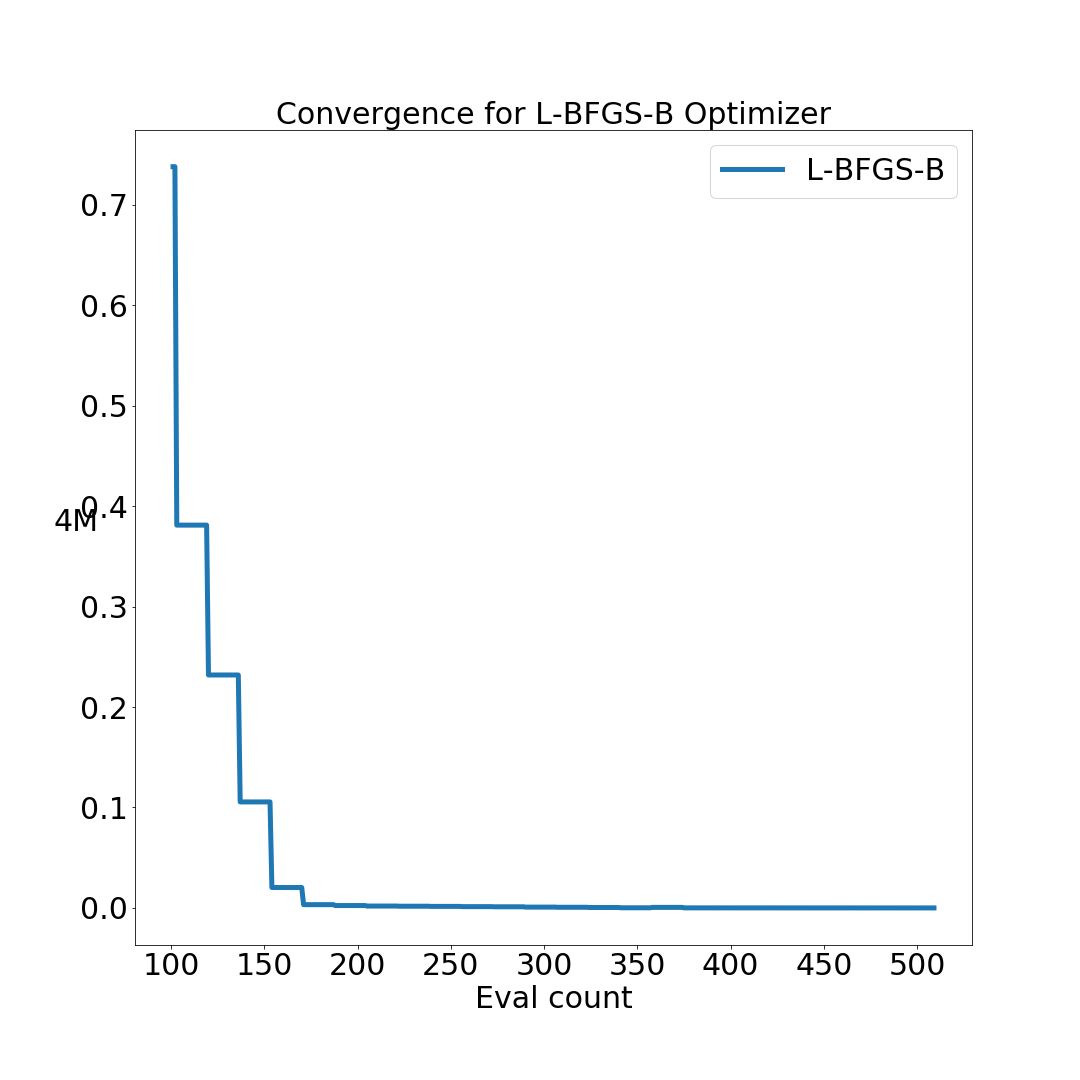}
  \caption{Convergence graph for the VQE computation of the Mass operator for Schwarzschild-deSitter black hole. The VQE result was obtained using the State-vector simulator with no noise and the Limited-memory BFGS Bound (L-BFGS-B)  optimizer.}
  \label{fig:Radion Potential}
\end{figure}
\begin{figure}
\centering
  \includegraphics[width = .4 \linewidth]{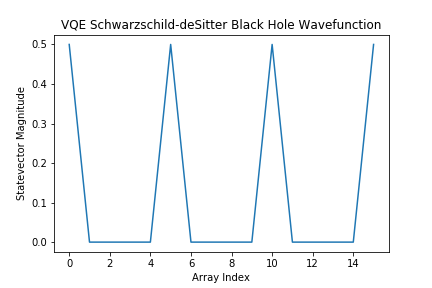}
  \caption{ VQE wave function for Schwarzschild-de Sitter black hole with $\lambda = .01$. Computations were done in four qubit representation with 16 states but only 4 of the  states satisfied the constraint ${ H}\left| \psi  \right\rangle  = 0$.  The Nariai limiting mass for $\lambda = .01$ is $4M_N = 13.333$.}
  \label{fig:Radion Potential}
\end{figure}
\begin{figure}[ht] 
  \label{ fig7} 
  \begin{minipage}[b]{0.5\linewidth}
    \centering
    \includegraphics[width=.75\linewidth]{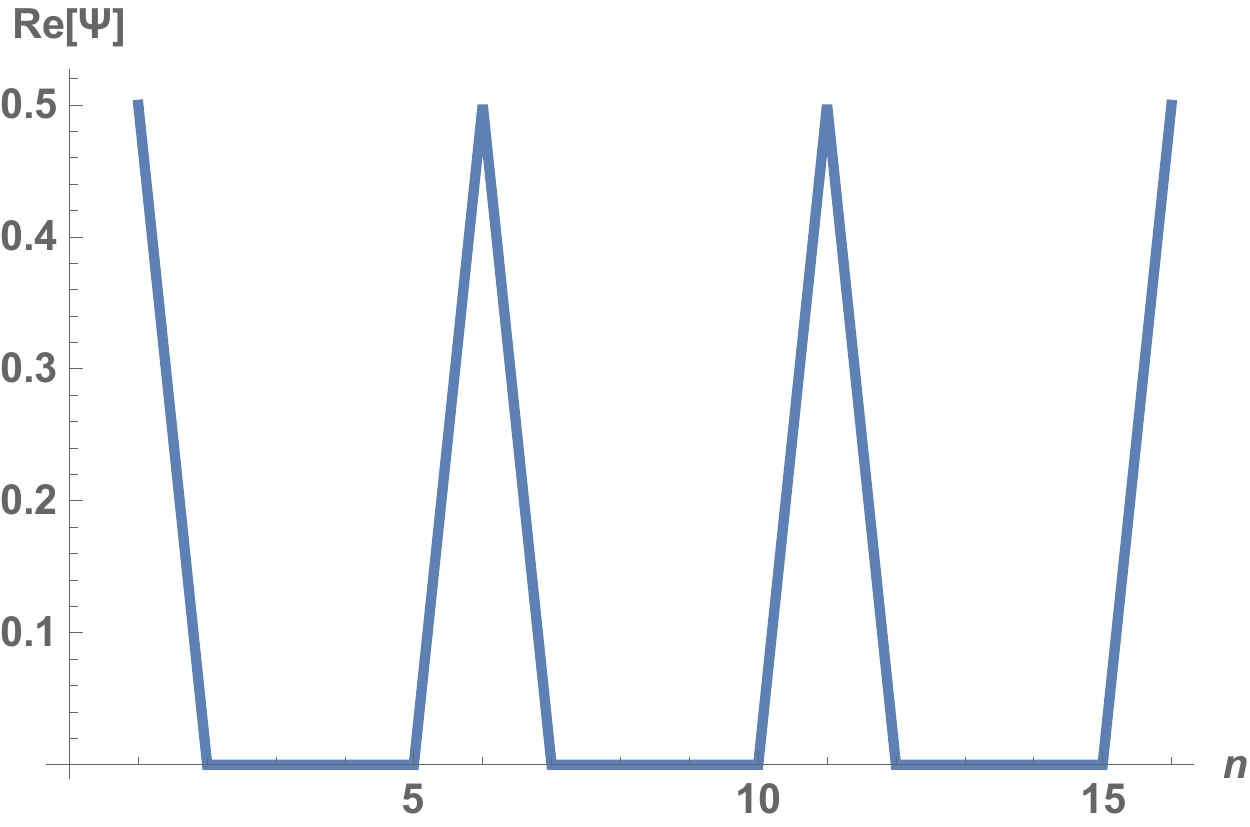}
  \end{minipage}
  \begin{minipage}[b]{0.5\linewidth}
    \centering
    \includegraphics[width=.75\linewidth]{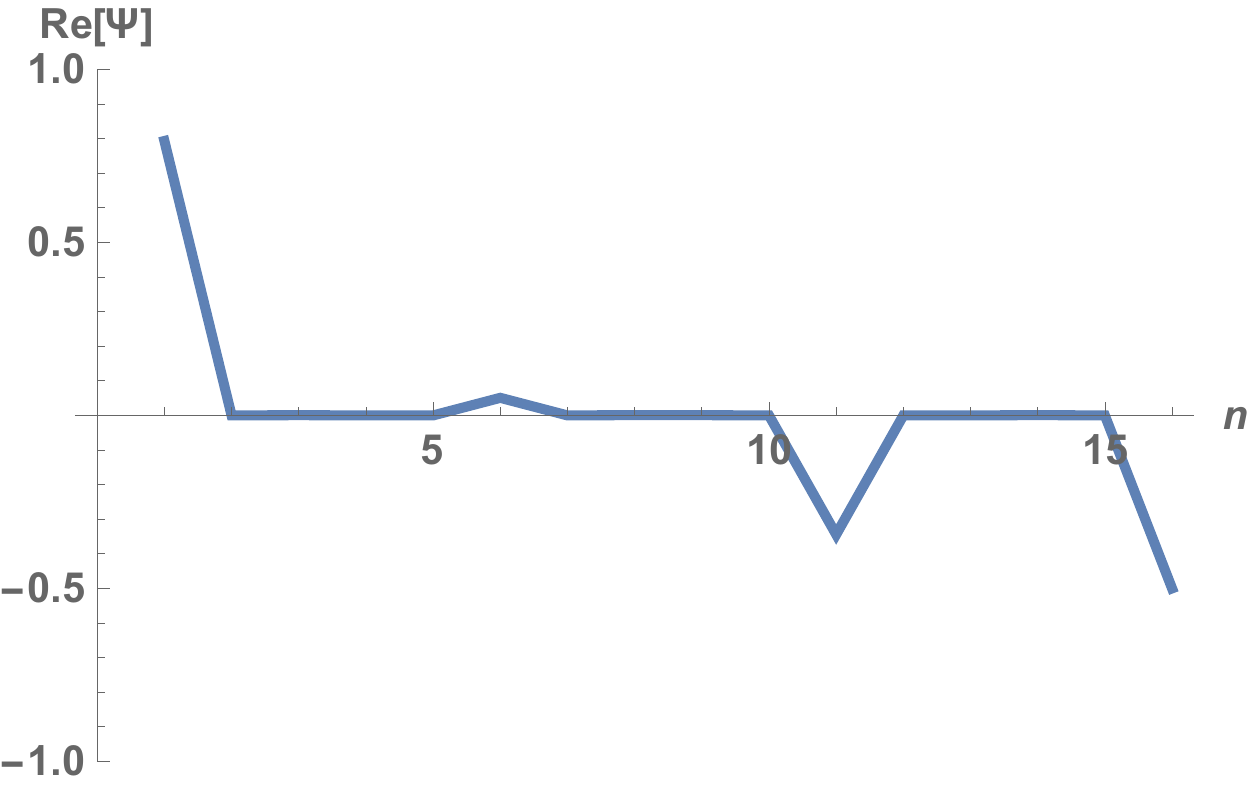} 
    \vspace{4ex}
  \end{minipage} 
  \begin{minipage}[b]{0.5\linewidth}
    \centering
    \includegraphics[width=.75\linewidth]{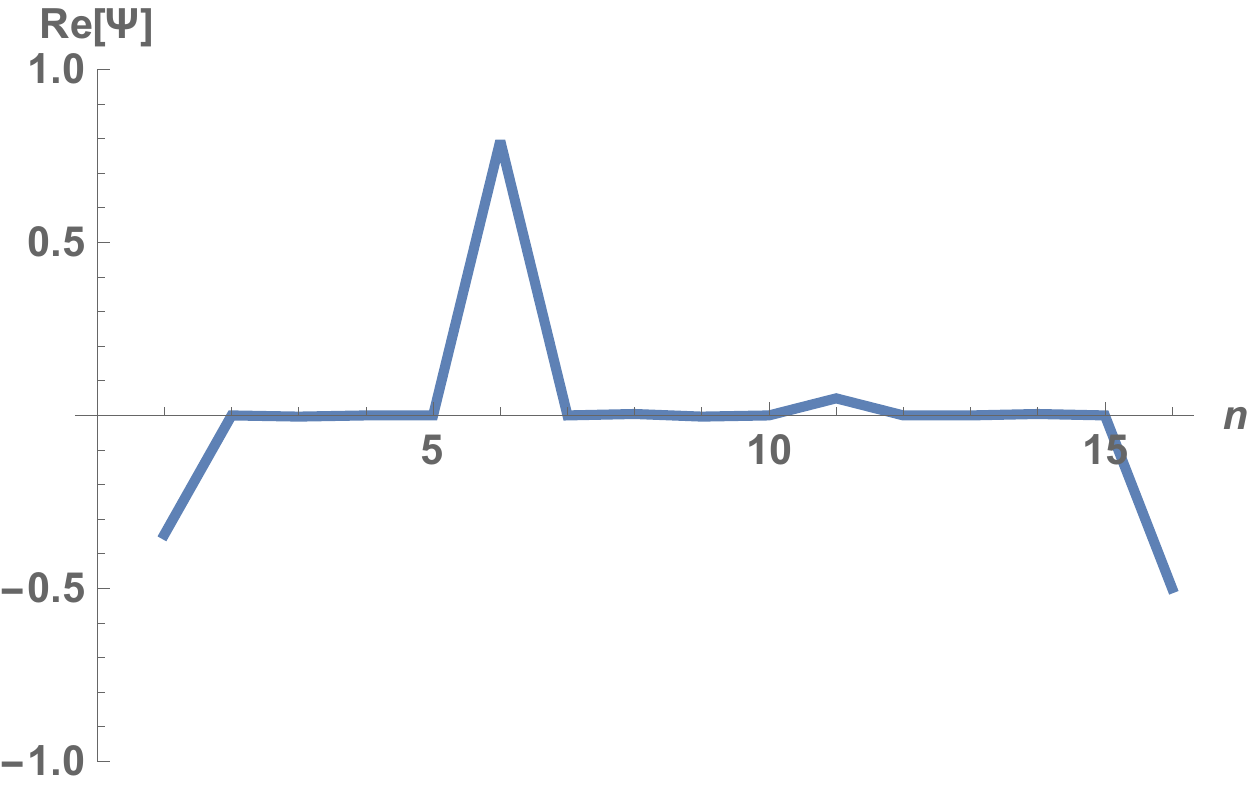} 
    \vspace{4ex}
  \end{minipage}
  \begin{minipage}[b]{0.5\linewidth}
    \centering
    \includegraphics[width=.75\linewidth]{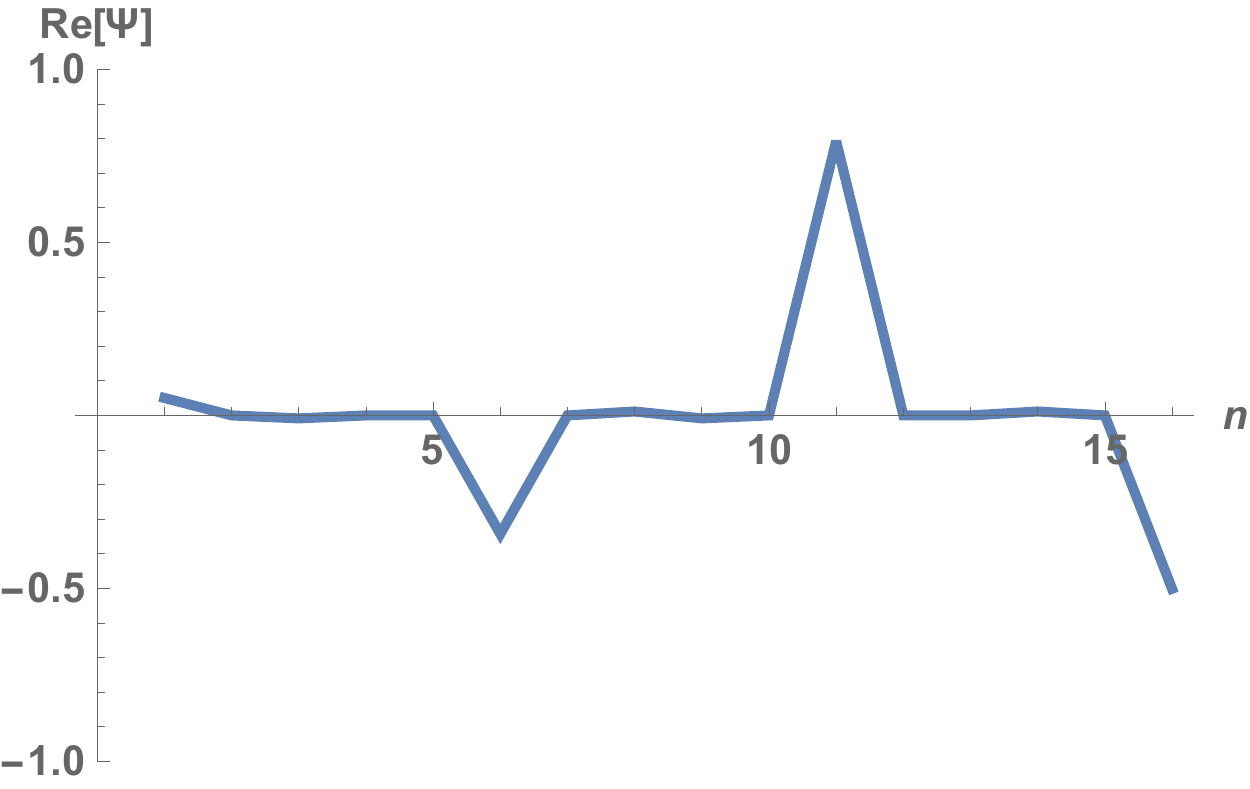} 
    \vspace{4ex}
  \end{minipage} 
    \caption{Wave functions of the Mass operator for Schwarzschild-de Sitter black hole for (upper left) with eigenvalues $4M =0$ and $\lambda = .01$, (upper right) $4M =0.935639$ and $\lambda = .01$, (lower left) $4M =3.29768,\lambda = .01$, (lower right) $4M =7.67034,\lambda = .01$. Computations were done in four qubit representation with 16 states but only 4 of the  states satisfied the constraint ${ H}\left| \psi  \right\rangle  = 0$.  The Nariai limiting mass for $\lambda = .01$ is $4M_N = 13.333$. }
\end{figure}
\newpage
\begin{figure}[htp] 
  \label{ fig7} 
  \begin{minipage}[b]{0.5\linewidth}
    \centering
    \includegraphics[width=.8\linewidth]{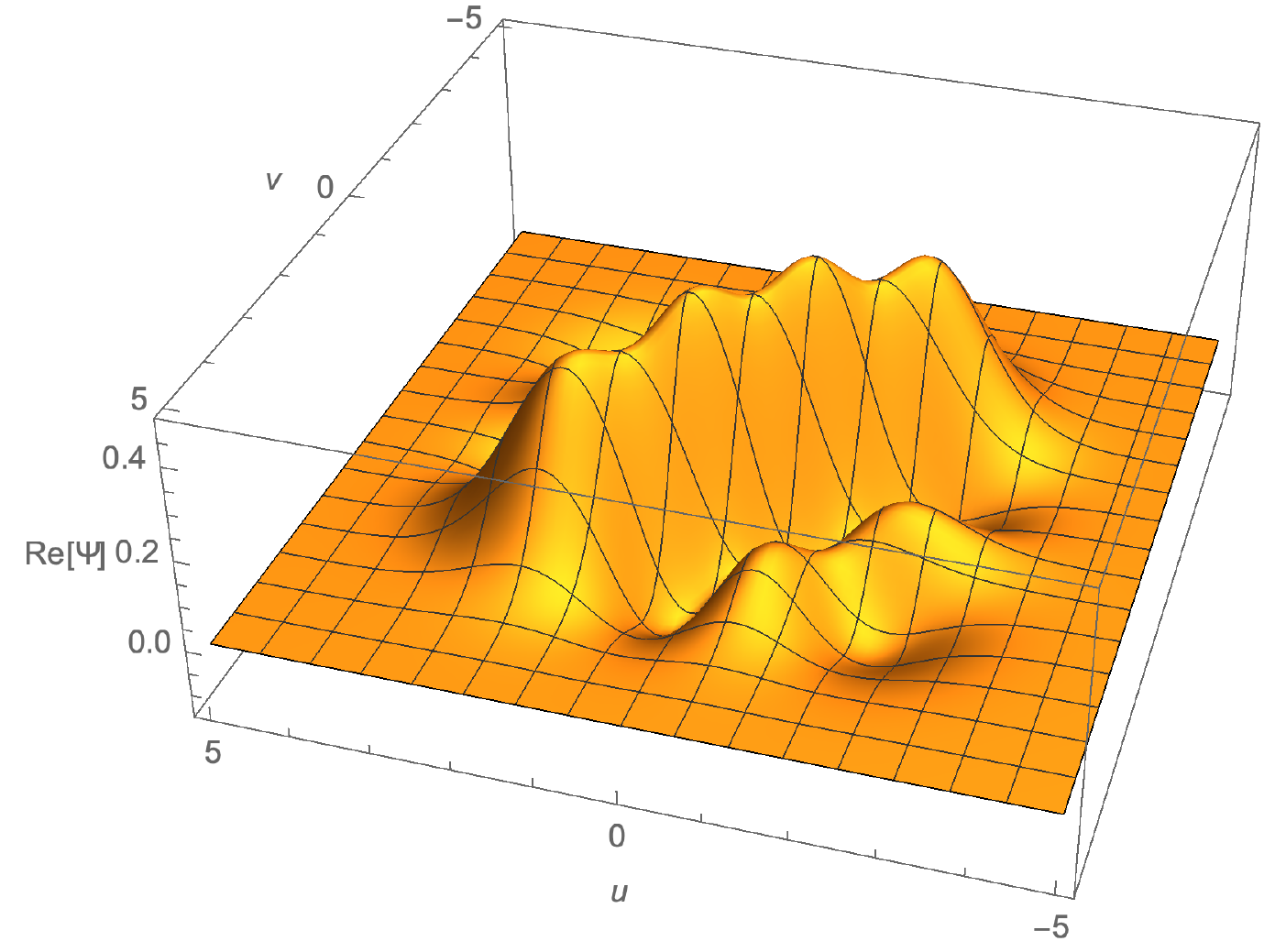}
    \vspace{4ex}
  \end{minipage}
  \begin{minipage}[b]{0.5\linewidth}
    \centering
    \includegraphics[width=.8\linewidth]{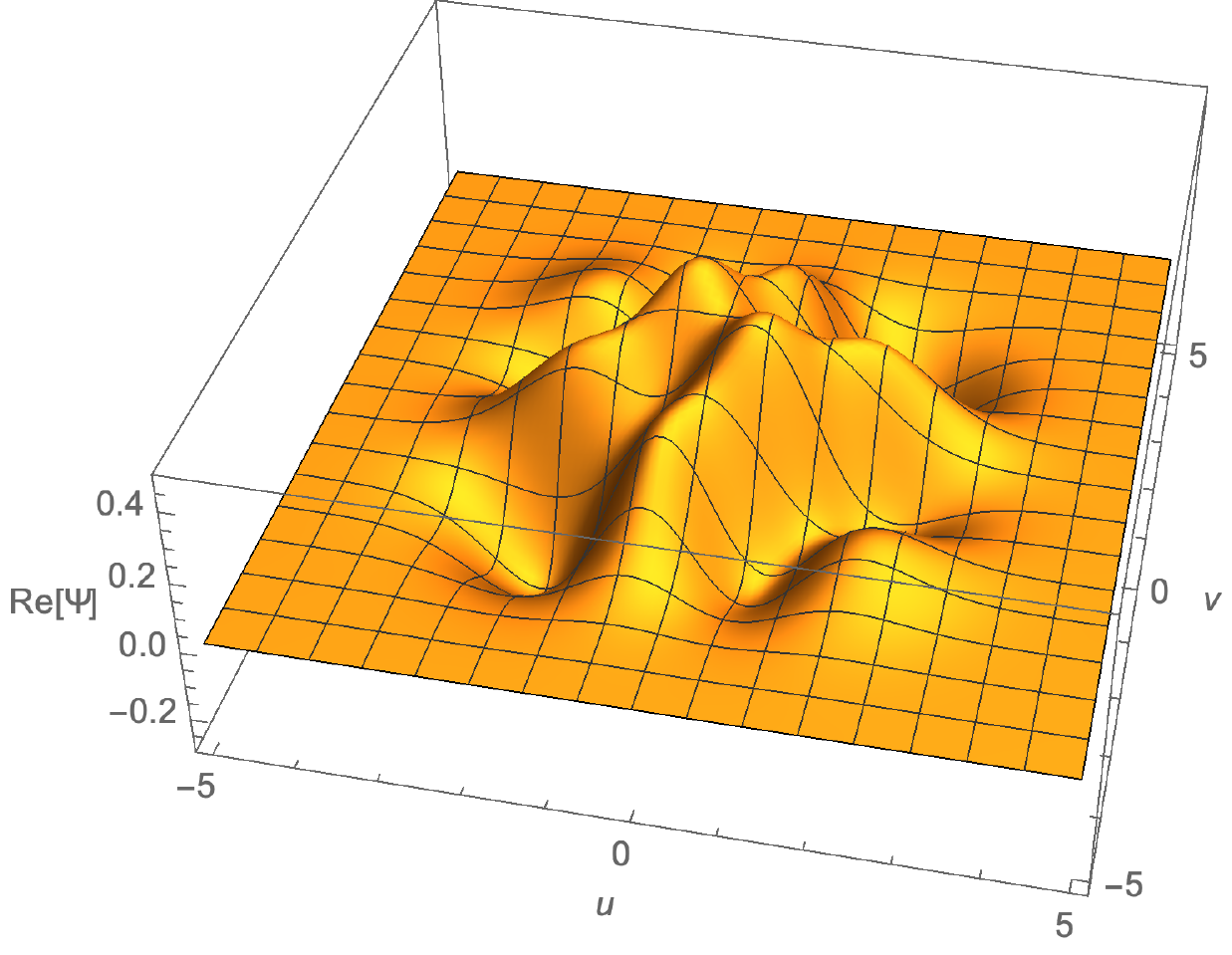} 
    \vspace{4ex}
  \end{minipage} 
  \begin{minipage}[b]{0.5\linewidth}
    \centering
    \includegraphics[width=.8\linewidth]{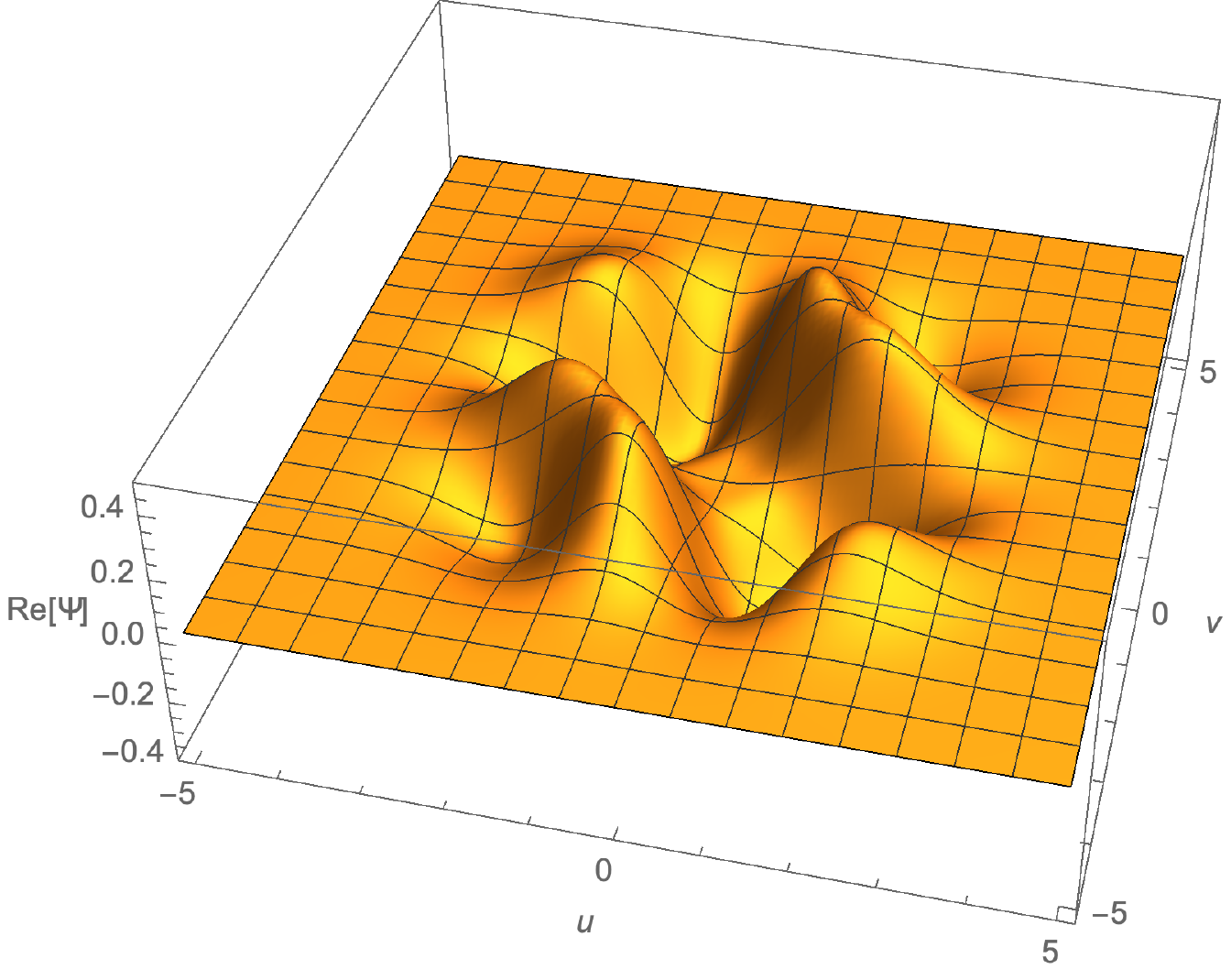} 
    \vspace{4ex}
  \end{minipage}
  \begin{minipage}[b]{0.5\linewidth}
    \centering
    \includegraphics[width=.8\linewidth]{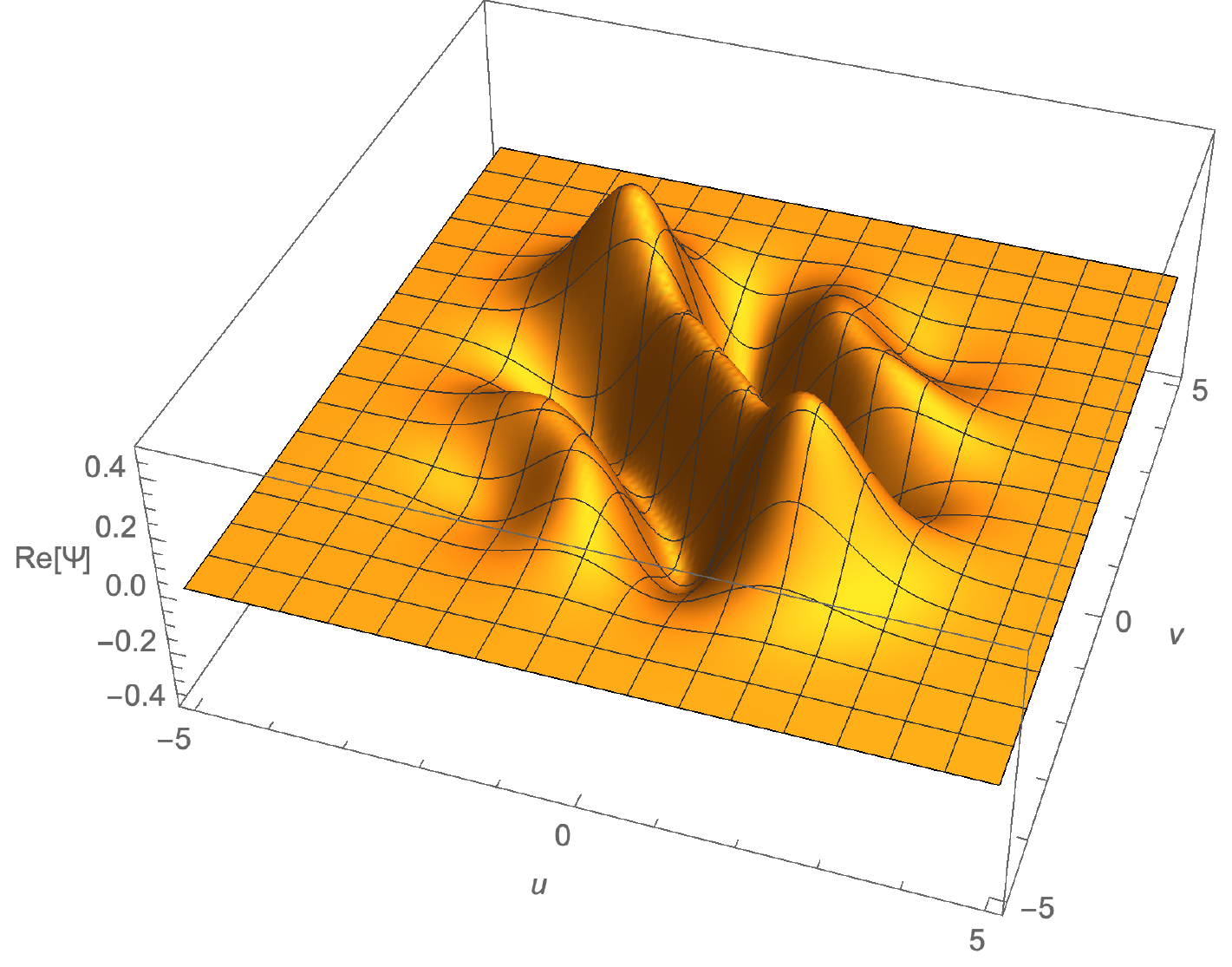} 
    \vspace{4ex}
  \end{minipage} 
    \caption{Wave functions of the Mass operator for Schwarzschild-de Sitter black hole using the harmonic oscillator wave function expansion for (upper left) with eigenvalues $4M =0$ and $\lambda = .01$, (upper right) $4M =0.935639$ and $\lambda = .01$, (lower left) $4M =3.29768,\lambda = .01$, (lower right) $4M =7.67034,\lambda = .01$.
    Computations were done in four qubit representation with 16 states but only 4 of the  states satisfied the constraint ${ H}\left| \psi  \right\rangle  = 0$. The Nariai limiting mass for $\lambda = .01$ is $4M_N = 13.333$.
    }
\end{figure}
\newpage
\begin{figure}[htp]
\centering
  \includegraphics[width = .8 \linewidth]{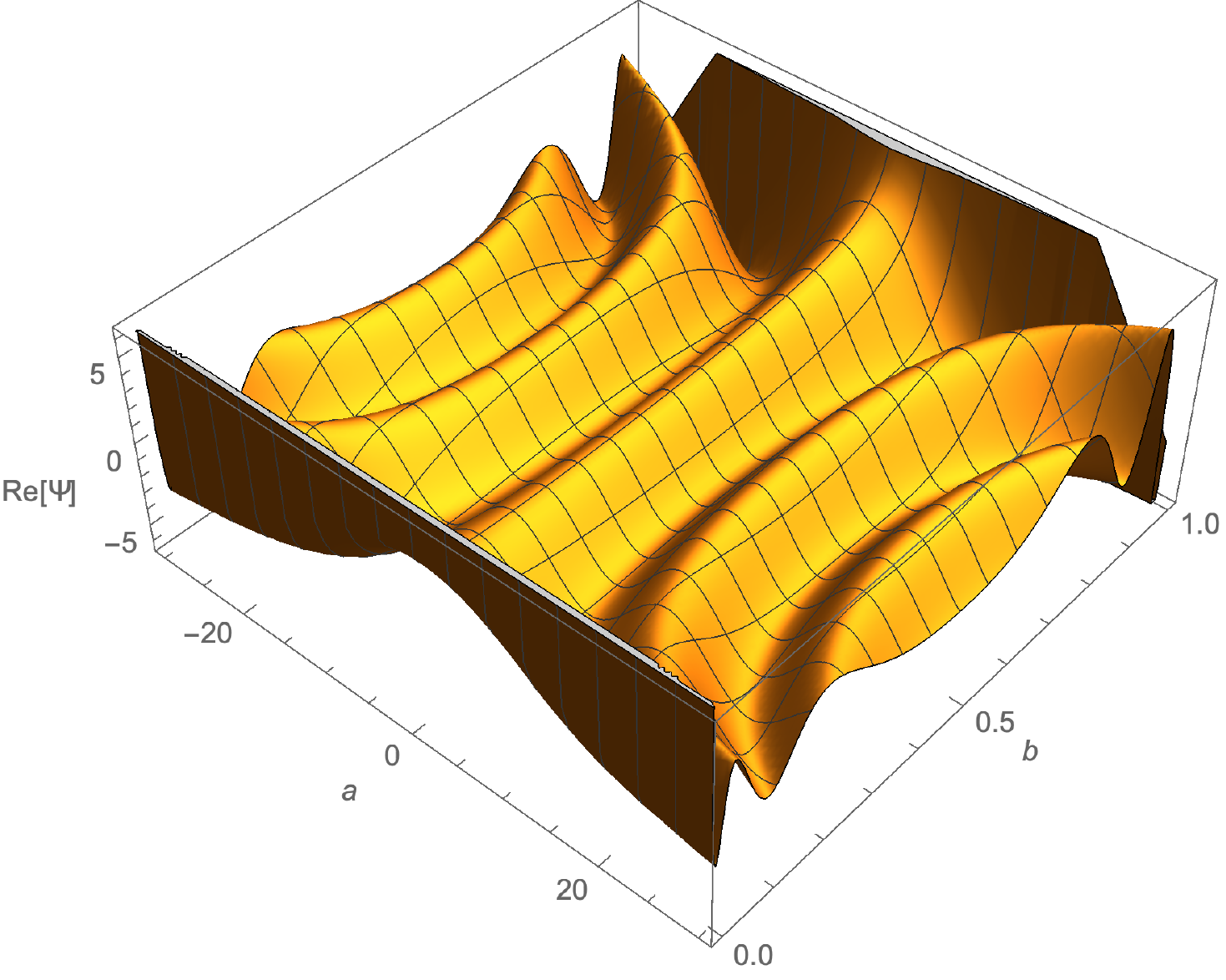}
  \caption{ WKB wave function for Schwarzschild-de Sitter black hole with $M=.5$ and $\lambda = .01$. The black hole horizon is located at $b =1.00337 $ and the cosmological horizon is located at $b=16.797$. The Narai limiting mass is for $\lambda= .01$ is $M_N= 3.333$.}
  \label{fig:Radion Potential}
\end{figure}
\newpage
The form of the quantum states of the black hole that is represented on the quantum computer are given by coefficients of a harmonic oscillator basis set in terms of a discrete vector basis and are shown in figure 11. To obtain a smooth description of the states  we can expand the solution for the mass eigenfunctions in terms of Hermite polynomials using these coefficients as:
\begin{equation}\Psi (u,v) = \sum\limits_{m,n} {{a_{m,n}}\frac{1}{{\sqrt {\pi {2^{m + n}}m!n!} }}{H_m}(u){e^{ - \frac{1}{2}{u^2}}}{H_m}(v){e^{ - \frac{1}{2}{v^2}}}} \end{equation}
Then the above list of coefficients for the basis vectors for the Hilbert space can be plotted as smooth functions of $u,v$ as shown figure 12. We see in figure 12 that the lowest mass state has the form of a squeezed state. In addition for large mass, a WKB-like wave function can give a good representation of a black hole wave function. This is given in \cite{Conradi:1994yy} in the $a,b$ minisuperspace coordinates as:
\begin{equation}{\Psi _{WKB}}(a,b) = \sqrt {\frac{{{b^{ - 3/2}}}}{{\frac{{\lambda {b^3}}}{3} + \frac{{2m}}{b} - 1}}} \text{exp}\left[ {iab\sqrt {\frac{{\lambda {b^3}}}{3} + \frac{{2m}}{b} - 1} } \right]\end{equation}
and is plotted in figure 13 for $m=.5 $ and $\lambda =.01$. It would be very interesting to compute the higher mass wave functions in quantum computing and compare with the WKB-like wave function of a large mass black hole. In addition there are extensions to the VQE algorithm to describe excited states \cite{vqe-excited}. These would be very interesting to apply to the Schwarzschild-de Sitter mass operator in order to determine  the accuracy of the masses that can be obtained through quantum computing and compare with the excited state masses and wave functions shown in figures 11 and 12.

\section{Conclusions}

In this paper we have considered the Schwarzschild-de Sitter black hole and Kantowski-Sachs cosmology is a completely quantum fashion by working with the Hamiltonian constraint and treating the mass as a eigenvalue of the mass operator. This may have advantages in closed cosmologies like Kantowski-Sachs or Schwarzschild-de Sitter which are not asymptotically flat. Quantum computing may also present an advantage because they can represent the path integral for the system in Lorentzian space directly and need not rotate into Euclidean space which might create ambiguities. To investigate this we have considered a simple minisuperspace approximation and used the VQE hybrid classical-quantum algorithm running on the IBM Qiskit state simulator to compute the lowest eigenvalue of the mass operator. We found very accurate results both for the lowest mass as well as for the lowest mass state wave function for the 4 qubit computation with  $16 \times 16$ matrices, but the results for 6 and 8 qubits using $64 \times 64$ and $256\times 256$ were less accurate. The structure of the quantum states for the Schwarzschild-de Sitter black hole we found using the qubit representation are quite interesting and deserve future investigations. In particular we would like to compute excited states of the mass operator to see what levels of accuracy can be obtained. For high mass levels we would like to investigate the approach to the Nariai limit and the representation of the WKB-like wave function of \cite{Conradi:1994yy}. It is  important to treat the inhomogeneous form of the metric that can be used to describe the formation of black holes, for example by the the collisions of spherical shells of high energy-density \cite{Gomez:1992fk}
\cite{Choptuik:1992jv}
\cite{Bak:1999wb}
\cite{Strominger:1993tt}. Finally the puzzling features of Schwarzschild-de Sitter black hole thermodynamics would be interesting to examine using quantum computing. One can can take an approach of thermodynamics using Hamiltonian quantum cosmology 
\cite{Cadoni:2021jer}
\cite{Gour:1999ta}
\cite{Gour:1999yu}
\cite{Giddings:2013vda}
\cite{Giddings:2021ipt} or the thermodynamics of Matrix models \cite{Susskind:2021dfc}. In either case quantum computing should be a useful tool to examine the states and thermodynamic observables \cite{Rinaldi:2021jbg}
\cite{Schaich:2022duk}
\cite{Gharibyan:2020bab}
\cite{Miceli:2019sym}.

\end{document}